\documentclass[subscriptcorrection,upint,varvw,barcolor=Goldenrod3,mathalfa=cal=euler,balance,hyphenate,french,pdf-a,nolists,table,xcdraw]{asmejour} %

\usepackage[switch]{lineno} 
\usepackage{lipsum}
\usepackage[noend]{algpseudocode}
\usepackage{algorithm}
\usepackage{algorithmicx}
\usepackage{etoolbox}

\usepackage{amsmath}

\usepackage{enumitem}

\DeclareMathOperator*{\argmax}{arg\,max}

\hypersetup{%
	pdfauthor={Saeid Bayat},                       		   	
	pdftitle={Multi-split configuration design for fluid-based thermal management systems},                  	
	pdfkeywords={Thermal Management System Design, Design Synthesis, Optimization, Optimal Flow Control, Knowledge Extraction, Graph Modeling},
	pdfsubject = {Describes the asmejour LaTeX template},			
	pdflicenseurl={https://ctan.org/pkg/asmejour},
}

\AtBeginEnvironment{algorithmic}{\lineskip0pt}

\usepackage{enumitem}
\usepackage{graphicx}
\usepackage{scalerel}
\setlist[enumerate,1]{label=\arabic*}
\setlist[enumerate,2]{label=\theenumi.\arabic*}
\setlist[enumerate,3]{label=\theenumii.\arabic*}

\JourName{Mechanical Design}

\begin{document}


\SetAuthorBlock{Saeid Bayat\CorrespondingAuthor}{Department of Industrial and Enterprise Systems Engineering,\\
   University of Illinois at Urbana-Champaign,\\
   Urbana, IL, USA \\
   email: bayat2@illinois.edu} 


\SetAuthorBlock{Nastaran Shahmansouri}{Autodesk Research,\\
   661 University Ave,\\
   Toronto, Canada \\
   email: nastaran.shahmansouri@autodesk.com} 
\SetAuthorBlock{Satya RT Peddada}{Department of Industrial and Enterprise Systems Engineering,\\
   University of Illinois at Urbana-Champaign,\\
   Urbana, IL, USA \\
   email: speddad2@illinois.edu} 
\SetAuthorBlock{Alex Tessier}{Autodesk Research,\\
   661 University Ave,\\
   Toronto, Canada \\
   email: alex.tessier@autodesk.com} 
\SetAuthorBlock{Adrian Butscher}{Autodesk Research,\\
   661 University Ave,\\
   Toronto, Canada \\
   email: adrian.butscher@autodesk.com} 
\SetAuthorBlock{James T Allison}{Department of Industrial and Enterprise Systems Engineering,\\
   University of Illinois at Urbana-Champaign,\\
   Urbana, IL, USA \\
   email: jtalliso@illinois.edu} 
   
\title{Advancing Fluid-Based Thermal Management Systems Design: Leveraging Graph Neural Networks for Graph Regression and Efficient Enumeration Reduction}


\begin{abstract}
In this research, we developed a graph-based framework to represent various aspects of  optimal thermal management system design, with the aim of rapidly and efficiently identifying optimal design candidates. Initially, the graph-based framework is utilized to generate diverse thermal management system architectures. The dynamics of these system architectures are modeled under various loading conditions, and an open-loop optimal controller is employed to determine each system's optimal performance. These modeled cases constitute the dataset, with the corresponding optimal performance values serving as the labels for the data. In the subsequent step, a Graph Neural Network (GNN) model is trained on 30\% of the labeled data to predict the systems' performance, effectively addressing a regression problem. Utilizing this trained model, we estimate the performance values for the remaining 70\% of the data, which serves as the test set. In the third step, the predicted performance values are employed to rank the test data, facilitating prioritized evaluation of the design scenarios. Specifically, a small subset of the test data with the highest estimated ranks undergoes evaluation via the open-loop optimal control solver. This targeted approach concentrates on evaluating higher-ranked designs identified by the GNN, replacing the exhaustive search (enumeration-based) of all design cases. The results demonstrate a significant average reduction of over 92\% in the number of system dynamic modeling and optimal control analyses required to identify optimal design scenarios.

\end{abstract}

\date{Version \versionno, \today}

\maketitle 


\section{Introduction}
\label{Sec: Introduction}
Prior knowledge is often leveraged to conceive new designs with enhanced performance. While this strategy can be effective for specific engineering tasks, it does present several shortcomings \cite{bayat2023extracting}. Firstly, this method falls short when there is a lack of existing design heritage knowledge, resulting in an initial design with potentially low performance that necessitates a lengthy process with multiple iterations for refinement. Secondly, the new design is intrinsically influenced by previous designs, and frequently, the design heritage itself is not optimal. As a result, the new design tends to mirror the design heritage closely, rendering the creation of radically new systems impossible. Finally, depending on design heritage to generate new designs is a time-consuming process. The knowledge accumulation happens over an extended period. A designer can acquire this knowledge either by gaining extensive experience in a specific domain or by undertaking the tedious and time-consuming task of reviewing numerous design scenarios. An alternate approach is required to address these shortcomings.

A design problem can be reframed as an optimization problem to address the first two challenges. In this scenario, an optimization problem describes specific goals as an objective function, which aims to maximize or minimize certain criteria. This optimization problem incorporates constraints, dynamics, initial conditions, and other pertinent factors. Once the problem setup is finalized, an optimizer is employed to solve the problem across various scenarios, generating valuable insights and data. This data effectively associates different scenarios with their respective objective function values. In the next step, the designer has the flexibility to select a design scenario that ensures optimal performance for the desired design task.

Although this method overcomes several shortcomings of the previous approach, the process of solving optimization problems can often be time-consuming. To address this issue, An Artificial Intelligence (AI) system, trained on a collection of optimal designs, can be deployed to unveil the relationships between different scenarios and their corresponding objective function values. This AI model can predict the objective value of the system under new scenarios, even those it was not trained on. This approach is capable of addressing the third challenge described in the first paragraph. This AI-based methodology offers three key advantages:

\begin{enumerate}
    \item It is suitable even for new designs that lack a design heritage.
    \item The knowledge created is based on optimal designs, guaranteeing that new designs informed by this knowledge are closely aligned with the optimal solution.
    \item The method is swift, capable of rapidly evaluating new design scenarios once the model is trained.
\end{enumerate}

This paper explores the design of active engineering systems utilizing the proposed AI-based methodology, with an emphasis on fluid-based thermal management systems. A set of optimal designs are required for training the AI model. To find the optimal performance of one system configuration, the dynamics of the system need to be modeled and optimized. The optimization problem setup necessitates the definition of four key components: the objective function, constraints, system dynamics, and controller \cite{bayat2023ss,bayat2023lgr}. The objective function measures the quality of the design. During the optimization process, the optimizer aims to maximize or minimize this function based on the system's objectives. The value of the objective function acts as a metric for assessing the design's performance. Constraints include the limitations or conditions that the optimization process must comply with. A design must meet all the outlined constraints to be deemed valid. Dynamics detail how the system changes over time in a specific scenario. Understanding dynamic behavior is vital for capturing the temporal aspects of system performance. Lastly, the controller, which significantly influences the system response, has parameters that are tuned to optimize the overall system performance. 

An automated framework can assist in generating a set of optimal systems. Such a framework should create a variety of configurations and set up and solve the optimization problem for each configuration. We have introduced an automated graph-based framework and methodology suitable for the class of thermal management systems in Ref.~\cite{bayat2023multi, bayat2023extracting}. The data used in this paper was generated using that framework. We will briefly describe how this works in the following sections. Graphs have applications in diverse domains, including circuit design \cite{herber2017advances}, vehicle suspension systems \cite{bayat2023control, herber2017enumeration}, and thermal management systems \cite{bayat2023multi, bayat2023extracting, peddada2019optimal}, making them an effective framework for this purpose. All system configurations can be systematically enumerated using a graph representation approach. An efficient enumeration technique developed by Herber et al.~\cite{herber2017advances} can streamline this process. In the subsequent step, the optimization problem needs to be derived and solved for each design configuration to determine its optimal performance. It is important to note that evaluating all of these cases is a time-consuming task, and one of the challenges that the proposed AI-based methodology aims to overcome.

In our previous work, we utilized various machine learning methods to extract interpretable knowledge for the thermal management system~\cite{bayat2023extracting}. While this approach provided valuable insights, aiding designers in gaining an initial understanding of the performance of various configurations, we encountered two issues. Firstly, with an increase in the number of system components, a new model had to be trained, necessitating the creation of new training data for each added component. Secondly, the machine learning models used in that study did not consider the graph structure, even though the connections among the components offer valuable insights for selecting the optimal configuration.

In this study, we utilize Graph Neural Network (GNN) to establish the mapping between design configurations, represented as graphs, and their respective objective function values. Deep neural networks have been employed in various applications to understand the relationship between a system's inputs and outputs~\cite{samek2021explaining}. However, a significant challenge emerges when the connectivity of components within the system is of great importance, as standard deep learning models are primarily based on Euclidean distance metrics \cite{zhou2020graph,wu2020comprehensive}. Expanding the application of deep neural models into non-Euclidean domains, often referred to as Geometric Deep Learning (GDL), has emerged as a new research area \cite{wu2020comprehensive}. Within this framework, there is a substantial emphasis on deep learning applications specifically designed for graphs~\cite{zhou2020graph,wu2020comprehensive}. This approach aims to minimize the need to investigate a multitude of cases to achieve optimal system design.

The structure of the subsequent sections in this paper is as follows: Section~\ref{Sec: Background} offers a comprehensive review of the background and related work, touching on graph theory, system representation, design optimization, and GNN. In Section~\ref{Sec: Methodology}, we explore the proposed methodology in detail, covering topics such as graph regression, Graph Attention Network (GAT)-GNN model, and metrics applied for model performance evaluation. Section \ref{Sec: HVAC}  presents a case study where the workflow setup specific to Fluid-Based Thermal Management System Optimization is discussed. Section ~\ref{Sec: Results} discusses the results, while Section ~\ref{Sec: lim} outlines the limitations of this work and provides suggestions for future research. The conclusion is presented in Section \ref{Sec: conc}.

\section{Background and Related Work}
\label{Sec: Background}

\subsection{Theory of Graphs}
\label{subSec: Graph_Theory}

A graph G is essentially a fundamental mathematical structure that can be expressed as a structured pair, denoted as (V, E). In this instance, V represents a set comprising various vertices or nodes that form the graph, while E denotes another set containing the edges, which essentially serve as links between these vertices. These edges are characterized by consisting of pairs of two elements chosen from the set V, demonstrating the relationships within the graph. The set of vertices, V, forms the building blocks of the graph, while the set of edges, E, presents a clear depiction of how these vertices are interconnected. This foundational concept underpins the understanding and analysis of complex networks, spanning from social structures to transportation systems, and beyond \cite{wu2020comprehensive}.

The graph adjacency matrix $A=(a_{{i,j}}){n\times n}$ can be defined as shown in Eq.~\eqref{eq: adj}. In unidirectional graphs, the matrix is symmetric, and all entries on the diagonal of matrix A are set to zero. This condition indicates that the graph does not contain any self-loop connections.  
\begin{align}
\label{eq: adj}
    a_{i,j} =
\begin{cases}
1, & \text{if } (v_i,v_j) \in  E \\
0, & \text{if } \mathrm{otherwise}
\end{cases}
\end{align}

\subsection{Graph-based Representation of Fluid Based Thermal Management Systems}
\label{subSec: repr_Hvac_Graph}

\begin{figure}[ht!]
    \centering
    \subcaptionbox{}{\includegraphics[width=1.0\linewidth]{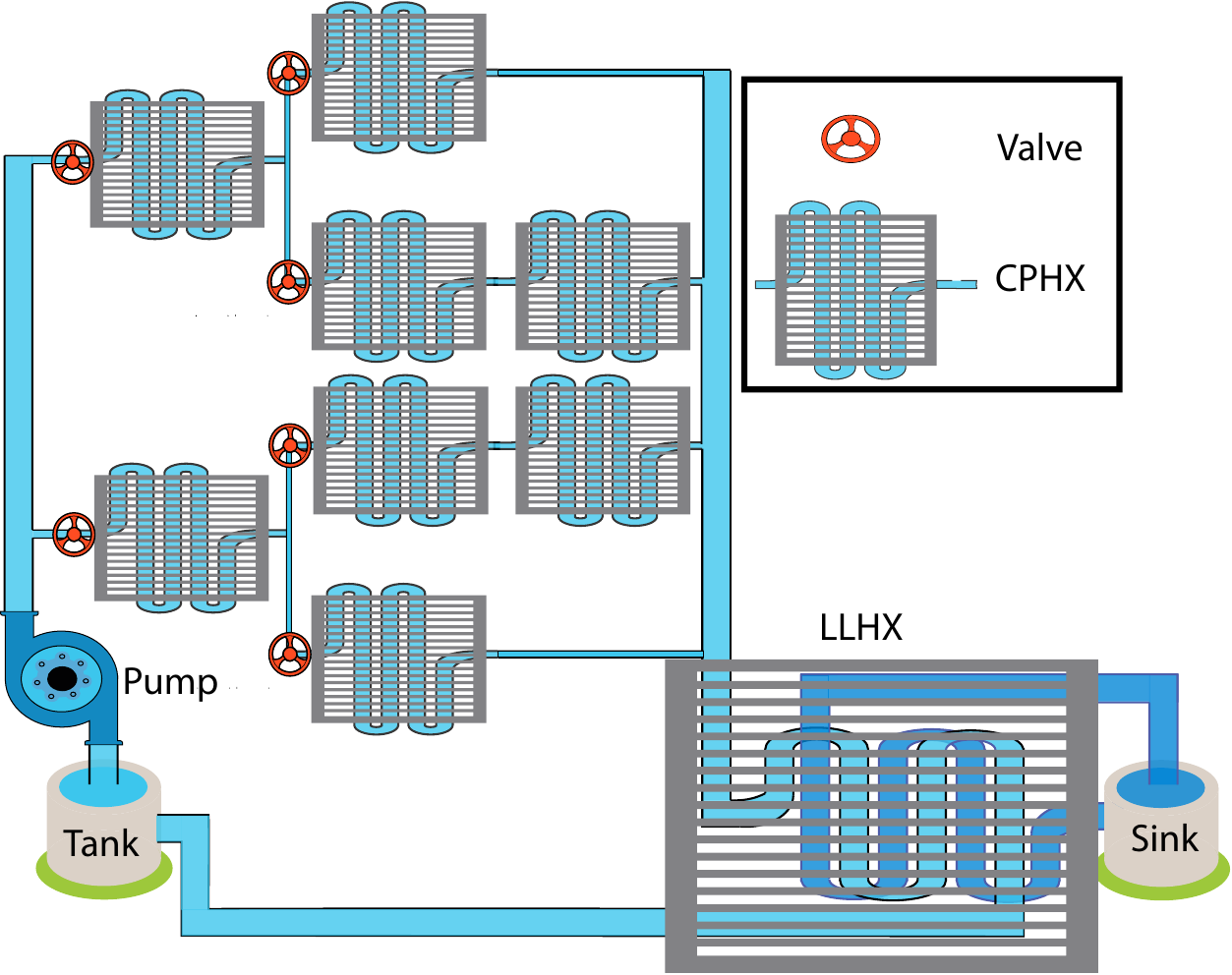}}
    \subcaptionbox{}{\includegraphics[width=1.0\linewidth]{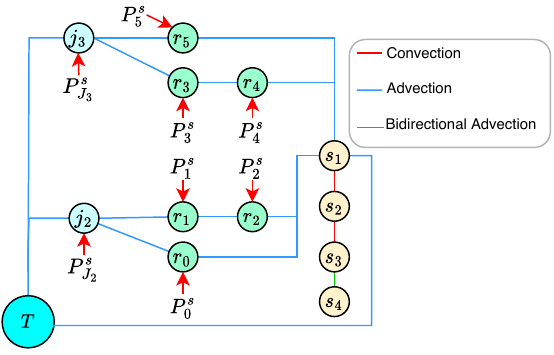}}
    \caption{ (a) Class of problems considered in this paper. The systems include a tank, a pump, valve(s), CPHXs in parallel and series, a LLHX, and a sink. (b) Graph modelling of the system.}
    \label{fig:whole_system}
\end{figure}

In this study, we examine a specific class of active systems, namely Fluid Based Thermal Management Systems. These systems include a tank, a pump, valve(s), various Cold Plate Heat Exchangers (CPHXs) arranged in parallel and series, a Liquid-to-Liquid Heat Exchnager (LLHX), and a sink. The primary objective of this system is to regulate the temperatures of various heat-generating devices attached to the CPHXs, through which a coolant circulates. The coolant, stored in a tank, is distributed to individual branches via a pump. Each branch, containing multiple valves, can distribute its incoming flow into sub-branches. As the coolant navigates the heat exchangers, it absorbs heat and subsequently transfers it to a thermal sink via a LLHX. The system's dynamics encompass advection, convection, and bidirectional advection ~\cite{bayat2023multi,peddada2019optimal}. Figure~\ref{fig:whole_system}(a) provides a simplified illustration, depicting a sample thermal management system with six CPHXs. It is worth noting that the system can comprise a more extensive network of nodes. Our investigation includes scenarios with a various number of CPHXs. 

Figure~\ref{fig:whole_system} (b) illustrates the graph representation of this system. In this depiction, $P$ represents the heat load applied to each node, and the dynamics between nodes are visually represented with distinct colors. The connections portray the fluid flow through the nodes. The system comprises four types of nodes: Tank ($T$), Junction nodes ($j_i$), regular nodes ($r_i$), and sink nodes ($s_i$). The vertices and connections for Tank, Junction nodes, and regular nodes can be defined based on this graph, as expressed in Eq.~\eqref{eq: HVAC_g}. 

\begin{align}
\label{eq: HVAC_g}
    V=\begin{bmatrix}
        T\\
        j_2\\
        j_3\\
        r_0\\
        r_1\\
        r_2\\
        r_3\\
        r_4\\
        r_5
    \end{bmatrix} \,\, , \, \,
    A=\begin{bmatrix}
        0 & 1 & 1 & 0 & 0 & 0 & 0 & 0 & 0\\
        1 & 0 & 0 & 1 & 1 & 0 & 0 & 0 & 0\\
        1 & 0 & 0 & 0 & 0 & 0 & 1 & 0 & 1\\
        0 & 1 & 0 & 0 & 0 & 0 & 0 & 0 & 0\\
        0 & 1 & 0 & 0 & 0 & 1 & 0 & 0 & 0\\
        0 & 1 & 0 & 0 & 0 & 1 & 0 & 0 & 0\\
        0 & 0 & 1 & 0 & 0 & 0 & 0 & 1 & 0\\
        0 & 0 & 0 & 0 & 0 & 0 & 1 & 0 & 0\\
        0 & 0 & 1 & 0 & 0 & 0 & 0 & 0 & 0
    \end{bmatrix}
\end{align}

The dynamics modeling of these systems as graphs has been established in a previous work~\cite{bayat2023multi}. This graph representation enables the formulation of dynamic equations for different nodes as Ordinary Differential Equations (ODEs). Furthermore, an optimal control problem has been defined with the objective of maximizing the thermal endurance of the system by controlling the fluid flow rate in pipes~\cite{bayat2023multi,bayat2023extracting}.

\subsection{Graph-Based System Generation and Evaluation}
\label{subSec: Graph_opt}

The graph optimization task involves finding the best graph, denoted as $G_{i^{\ast}}$, among a set of graphs $G_1, G_2, …, G_N$, that maximizes an objective value function $J(G_i)$. To identify the optimal graph, two methods are required: 1) a method to construct the graph $G_i$, and 2) a method to calculate its corresponding objective value $J(G_i)$. The computational time needed for these steps can vary depending on the specific system under consideration.

According to Ref.~\cite{sirico2023use,guo2018reducing}, the computation time of these steps can lead to three types of scenarios:

Type 0: In this scenario, all graphs can be generated and evaluated within the allocated time, denoted as T.
Type 1: Here, all graphs can be generated, but only a subset of them can be evaluated within the allocated time T.
Type 2: This scenario arises when not all graphs can be generated within the allocated time T.
In the above classification, T represents the time allocated for completing the graph design study.

In Type 0 scenarios, enumeration can be employed to identify the optimal graph. This means that all possible graphs are generated and evaluated, allowing for the selection of the graph with the highest objective value. In Type 1 scenarios, although all graphs can be enumerated, only a subset of them can be evaluated within the allocated time. In such cases, machine learning tools can be utilized to establish a mapping between the generated graphs and their corresponding objective values. By training a model on the subset of evaluated graphs, it becomes possible to estimate the objective values for all graphs in this category. This trained model can then be used to predict the objective values for the remaining graphs.

In Type 2 scenarios, where not all graphs can be generated within the allocated time, two potential approaches can be considered. The first approach involves training a Graph Neural Network (GNN) based on the subset of generated graphs. This trained model can then be used to generate new graphs, and the same evaluation method used in Type 1 can be applied to evaluate these generated graphs \cite{sirico2023use}. The second approach in Type 2 scenarios involves utilizing the latent space obtained through Generative Adversarial Networks (GANs). The graphs are optimized based on this latent space, which is a concept explored in topology optimization. Various methods such as encoder-decoder, active learning, genetic algorithms, and more have been utilized in the literature for graph generation in this category \cite{guo2018indirect,chen2019synthesizing}.

Active Engineering systems typically fall into the categories of Type 1 and Type 2. Optimizing and evaluating these active systems requires optimal control, which can be computationally expensive. Furthermore, as the number of components in a system increases, the graph generation process becomes more complex, leading to a combinatorial explosion. This makes it infeasible to generate and evaluate all possible graphs. To address this challenge, specific methods need to be developed for Type 1 and Type 2 scenarios, enabling the design of high-performance systems. These methods allow for the generation and evaluation of a subset of graphs within the allocated time, while still providing valuable insights. This work focuses on type 1.

\subsection{Design Optimization of Active Engineering System}
\label{subsec: OLOC_formula}

Dynamic engineering systems typically include an active module known as a controller, which can be adjusted to optimize the system's objective function while considering dynamics, constraints, and other factors. The controller significantly impacts the system's response and falls into two broad categories: open-loop control and closed-loop control \cite{bayat2023nested}. Open-loop control assumes no predefined structure for the controller, allowing it to adapt freely (while adhering to all design constraints) to optimize the objective function value. This controller possesses complete information over the entire time horizon, which is often unfeasible in practice. On the other hand, closed-loop control maintains a fixed structure and possesses limited information compared to open-loop control. Nevertheless, it is implementable in real-world scenarios because it employs feedback from the system to adjust its response and can be used to demonstrate the system's stability.

Utilizing open-loop control in the initial phases of system design is preferable because it allows us to harness the maximum potential of the system without confining it to a fixed closed-loop controller \cite{allison2014co}. Subsequently, in the later stages, a realizable closed-loop controller can be developed. Optimization of open-loop control can be achieved through two main categories: direct and indirect methods \cite{allison2013multidisciplinary,bayat2023ss,bayat2023lgr}. In the indirect method, the Pontryagin maximum principle and Lagrange equations are employed to derive optimality equations, followed by the use of numerical methods to solve the optimization problem. Conversely, in the direct method, the problem is discretized and transformed into a nonlinear program, and powerful NLP optimizers such as SNOPT~\cite{gill2005snopt} and IPOPT~\cite{biegler2009large} are employed. Direct methods are considerably more efficient than indirect methods \cite{bayat2023ss}, and software tools like GPOPS~\cite{patterson2014gpops} and Dymos~\cite{falck2021dymos} can be utilized to solve these problems.

The general Open-Loop Optimal Control (OLOC) problem formulation is represented by Eq.~\eqref{eq: Ol_formula} \citep{bayat2023lgr}. In this equation, various variables are defined as follows: $t$ denotes time, $t_0$ is the initial time, $t_f$ is the final time, $u$ represents the control, $\bm \xi$ signifies the state, $\Phi$ corresponds to the Mayer cost, $\mathcal{L}$ pertains to the Lagrange cost, $\mathcal{\bm C}$ denotes path constraints, and $\mathcal{\bm \phi}$ represents boundary constraints. The dynamics of the system are described by $\bm f$. It is worth noting that the notation for the dynamics in the OLOC formula may vary depending on the method employed \cite{bayat2023lgr,bayat2023ss}, but it will not explored further in this paper because of brevity.

\begin{align}
    \label{eq: Ol_formula}
    \min_{\bm \xi(t) \, , \, u(t)}\,\,\, &J=\, \Phi\left( \bm \xi(t_0), t_0, \bm \xi(t_f), t_f \right)+ \int^{t_f}_{t_0} \mathcal{L}\left(t,\bm\xi(t),\bm u(t) \right)dt \\
    &\mathrm{Subject\,\, to:}\nonumber \\
    &\dot{\bm \xi}-\bm f\left(t,\bm \xi(t),\bm u(t)\right)=0\nonumber \\
    &\mathcal{\bm C}\left( t,\bm\xi(t),\bm u(t) \right)\le0\nonumber \\
    &\mathcal{\bm \phi} \left(t_0,\bm \xi(t_0),t_f,\bm \xi(t_f) \right)= 0\nonumber 
\end{align}

When we have a graph model denoted as $G_i$, its associated dynamical system is referred to as the "system dynamics," which is represented by the equation $\bm f$. Given this graph, OLOC solver aims to determine the optimal state and controller for the system. The objective is to minimize the objective function $J$. Consequently, the corresponding optimal objective value function for system $i$ with graph modeling $G_i$ is denoted as $J(G_i)$. This study utilizes OLOC using direct method. A comprehensive overview of the formulation of the optimal control problem in fluid-based thermal management systems, as considered in this study, is presented in our previous work~\cite{bayat2023multi}.

\subsection{Message Passing Neural Network: GAT-GNN}
\label{subSec: GNN}
With the significant advancements in various machine learning methods and their high performance across different applications, they have found applications in handling graph data. Initially, the literature explored recursive and recurrent neural networks for this purpose \cite{wu2020comprehensive,zhou2020graph}. These networks learn the representation of a target node by iteratively propagating information from its neighbors until a stable fixed point is achieved. However, this process is computationally intensive, and recent efforts have focused on addressing these computational challenges.

Drawing inspiration from the successful application of Convolutional Neural Networks (CNNs) in computer vision, researchers have developed numerous methods that redefine the concept of convolution specifically for graph data. CNNs are known for their ability to extract multi-scale localized spatial features and combine them to create highly expressive representations. However, it is important to note that CNNs were originally designed for regular Euclidean data, such as images (2D grids) and texts (1D sequences) \cite{wu2020comprehensive}. In contrast, graph data structures are non-Euclidean in nature. Graphs are inherently sparse and exhibit irregularities, with varying numbers of unordered nodes and neighbors. As a result, applying operations like convolutions in the graph domain is more complex and less straightforward compared to the image domain.

In addition, it is worth noting that existing machine learning algorithms typically make the assumption that instances are independent of one another \cite{zhou2020graph}. However, this assumption does not hold for graph data, as each instance (node) is connected to others through various types of links. As a result, the extension of deep neural models to non-Euclidean domains, commonly referred to as Graph Deep Learning (GDL), has emerged as a rapidly growing research area. Within this framework, applying deep learning techniques to graphs has gained significant attention and interest.

The core concept underlying Graph Neural Networks (GNNs) is the process of message passing, which involves the exchange of information between nodes and their neighbors across multiple layers. Additionally, the edges connecting these nodes can have different weights to indicate the significance of each connection. Various GNN models differ in how they handle the propagation of these messages between nodes across different layers. Message Passing Neural Networks (MPNNs) provide a structured approach to understand and design GNNs by defining how information is exchanged and aggregated across nodes in a graph \cite{gilmer2020message}. Other GNN architectures, such as GraphSAGE \cite{hamilton2017inductive}, Graph Attention Networks (GAT) \cite{veličković2018graph}, and Graph Isomorphism Networks (GIN) \cite{xu2019powerful}, can be viewed as specific instances or variations within the MPNN framework.

Consider a graph $G = (V,E)$, comprising a node set $V$ and an edge set $E$. For a given node $u \in V$, let $N_u$ denote its neighborhood. Furthermore, let $\mathbf{x}_u$ represent the features of node $u \in V$, and $\mathbf{e}_{uv}$ denote the features of the edge $(u, v) \in E$. The formulation of a Message Passing Neural Network (MPNN) layer can be articulated as follows \cite{bronstein2021geometric}:

\begin{align}
    \mathbf{h}_u = \phi \left( \mathbf{x}_u, \bigoplus_{v \in N_u} \psi(\mathbf{x}_u, \mathbf{x}_v, \mathbf{e}_{uv}) \right)
\end{align}

In the context of an MPNN computational block, the representations of graph nodes are updated by aggregating messages received from their neighbors. This process involves the use of two differentiable functions, denoted as $\phi$ (the 'update' function) and $\psi$ (the 'message' function). Here, $\mathbf{h}_u$ represents the node representation for node $u$. The update function $\phi$ is responsible for incorporating the aggregated messages from neighbors and updating the node representation. The message function $\psi$ defines the transformation applied to the node and its neighbors to generate the messages exchanged between them. To aggregate the messages from neighbors, a permutation-invariant aggregator operator $\bigoplus$ is used. This operator can handle various types of inputs, such as element-wise sum, mean, or max, and ensures that the aggregation process is insensitive to the order in which the messages are received.

The outputs of one or more MPNN layers are node representations, denoted as $\mathbf{h}_u$, for each node $u \in V$ in the graph. These node representations serve as valuable inputs for various downstream tasks, including node classification, graph classification, regression, or edge prediction. By leveraging the node representations generated by MPNNs, one can effectively tackle a wide range of tasks involving graph-structured data.

In this work, Graph attention network is utilized.
A multi-head GAT layer can be expressed as follows \cite{veličković2018graph}:
\begin{align}
\label{eq: GAT}
\mathbf{h}_u &= \overset{K}{\underset{k=1}{\Big\Vert}} \sigma \left(\sum_{v \in N_u} \alpha_{uv} \mathbf{W}^k \mathbf{x}_v\right)\\
\alpha_{uv} &= \frac{\exp(\text{LeakyReLU}(\mathbf{a}^T [\mathbf{W} \mathbf{h}_u \Vert \mathbf{W} \mathbf{h}_v \Vert \mathbf{e}_{uv}]))}{\sum_{z \in N_u}\exp(\text{LeakyReLU}(\mathbf{a}^T [\mathbf{W} \mathbf{h}_u \Vert \mathbf{W} \mathbf{h}_z \Vert \mathbf{e}_{uz}]))}
\end{align}

Here, $K$ represents the number of attention heads, $\Big\Vert$ denotes vector concatenation, $\sigma(\cdot)$ serves as an activation function (e.g., ReLU), $\alpha_{ij}$ signifies attention coefficients, and $W^k$ stands for a matrix of trainable parameters for the $k$-th attention head. Additionally, $\mathbf{a}$ constitutes a vector of learnable weights, $\cdot^T$ denotes transposition, and $\text{LeakyReLU}$ denotes a modified ReLU activation function. The attention coefficients undergo normalization to facilitate easy comparison across different nodes \cite{veličković2018graph}. In this context, the attention coefficient $\alpha_{uv}$ gauges the "importance" of node $u \in V$ to node $v \in V$.

\section{Methodology}
\label{Sec: Methodology}

The methodology includes four sub-sections. We review the GNN-GAT model employed in the work in Sec.~\ref{Sec: GNN-GAT}. Next, Sec.~\ref{Sec: Graph-Regression} briefly discusses why a graph regression method has been employed. The metrics and evaluation functions used for the analysis and comparisons of the results are described in Sec.~\ref{Sec: metrics}. Lastly, a description of the data set required for the analysis is presented in Sec.~\ref{Sec: data}.

\subsection{GNN-GAT Model}
\label{Sec: GNN-GAT}
This research investigates graph regression using GNN. The GNN receives input features from all nodes in the graph, including the heat loads and node types, as well as information about the connections between the nodes in the form of an adjacency matrix. The GNN then generates a continuous variable as the output of the graph, which represents the estimated objective value for the corresponding graph. By utilizing available training data and employing Mean Squared Error (MSE) loss, the GNN adjusts its parameters to closely align the output (estimated objective value) with the target values (true objective values). The GNN architecture in this study consists of four layers:

\begin{enumerate}
    \item The initial layer consists of three multihead Graph Attention Network (GAT) layers. These GAT layers extract features from nodes and generate new features based on Eq.~\eqref{eq: GAT}
    \item Subsequently, mean layers are employed to compute the average over all heads.
    \item To transform node features into graph features, three pooling layers are employed, including global mean, global max, and global min pooling.
    \item The final layer is a linear layer mapping the graph embedding to a single continuous variable.
\end{enumerate}

\subsection{Graph Regression}
\label{Sec: Graph-Regression}

GNN can be used for graph regression or graph classification and the choice of graph regression over classification is often driven by the need for a more detailed understanding of the relationships between graph features and their corresponding target values. Graph regression is particularly valuable when the goal is to obtain specific insights into the performance or other continuous characteristics of graphs, rather than simply categorizing them into discrete classes. In the study by Sirico et al.~\cite{sirico2023use}, graph classification is implemented to categorize data into two groups: "good" and "bad." This classification is based on the median performance value, dividing graphs into binary classes. While this approach can be useful for certain applications, it lacks detailed insights into the quality of the graphs. For example, if 30 out of 80 graphs are identified as "good," their relative ranking remains unknown. To address this limitation, this paper employs graph regression to provide a more nuanced evaluation of graph quality.

Graph regression involves predicting a continuous value or set of values for a given graph based on its features. In the context of supervised graph regression, we work with a dataset of graphs $G_i$, each associated with a continuous target variable, often denoted as $J(Gi)$. The goal is to train a model using various graph properties, such as structural attributes, embeddings, and node-level information, to accurately predict the target variable. This approach is particularly useful when the objective is to predict performance metrics, numerical scores, or other continuous measures associated with graphs. For example, in a scenario where the performance of different graphs is measured on a continuous scale; the GNN model aims to generalize from the training data to make accurate predictions of performance for new, unseen graphs.

\subsection{Metrics and Evaluation Function}
\label{Sec: metrics}
This section describes the parameters, metrics, and evaluation functions of this study which have been carefully chosen to help analyzing and evaluating different aspects of the system under investigation. Here, $G_i$ denotes the ith graph, $J(G_i)$ represents the true objective value of the graph $G_i$, $\hat{J}(G_i)$ indicates the predicted objective value of the graph $G_i$. The index $i^{\ast}$ corresponds to the graph index that yields the highest objective value and can be obtained as $i^{\ast} = \argmax_i J(G_i)$, with its corresponding objective value denoted as $J^{\ast} = J(G_{i^{\ast}})$. Similarly, $\hat{i}^{\ast}$ is the graph index that yields the highest predicted objective value, obtained as $\hat{i}^{\ast} = \argmax_i \hat{J}(G_i)$, and its corresponding objective value is $\hat{J}^{\ast} = \hat{J}(G_{\hat{i}^{\ast}})$. Furthermore, $J(G_{\hat{i}^{\ast}})$ represents the true objective value of the predicted optimal graph, and $\hat{J}(G_{i^{\ast}})$ represents the predicted objective value of the true optimal graph.

In graph regression, the evaluation of model performance relies on metrics that assess the accuracy and reliability of the predicted continuous values compared to the actual target values. Several metrics commonly used in regression analysis can be adapted for graph regression scenarios \cite{botchkarev2018performance}. 

Some key metrics often employed in the context of graph regression include MSE, Mean Absolute Error (MAE), R-squared ($R^2$), and Root Mean Squared Error (RMSE). These metrics offer a comprehensive evaluation of the model's predictive performance in graph regression, allowing researchers and practitioners to assess the accuracy and reliability of the model's predictions. The choice of metrics depends on the specific characteristics and requirements of the graph regression task at hand. In this study, we employed MSE to adjust GNN parameters.

    As shown in Eq.~\eqref{eq: mse}, MSE measures the average squared difference between predicted and actual values. It is a widely used metric for regression tasks, providing a measure of the overall model accuracy. Here, \(n\) is the number of samples, \(J(G_i)\) is the true target value for the \(i\)-th graph, and \(\hat{J}(G_i)\) is the predicted value.
    \begin{align}
    \label{eq: mse}
        \text{MSE} = \frac{1}{n} \sum_{i=1}^{n} (J(G_i) - \hat{J}(G_i))^2
    \end{align}

While the above mentioned metrics offer a solid measure of the regression problem's performance, they might not accurately reflect the objective of our study. Our aim is to minimize the enumeration to a subset with a smaller population, where the members in this subset demonstrate superior performance compared to the larger population. Consequently, prioritizing the ranking of these members becomes a crucial criterion to demonstrate the effectiveness of the trained GNN. In light of this, the subsequent evaluation functions can be explored for this task.

\subsubsection{Kendall Tau Rank Distance } 
\label{Metric: k}

    Kendall's tau \cite{sen1968estimates} (K) is a statistical measure employed to assess the degree of similarity or agreement between two rankings. This metric is particularly useful when dealing with data that is ranked or ordered, such as preferences, or performances. The basic idea behind Kendall's tau is to compare the pairwise \emph{concordant} and \emph{discordant} pairs of elements between two rankings.

In the context of Kendall's tau, a "concordant pair" refers to two pairs of elements that share the same order in both rankings, meaning that if one element is ranked higher than another in the first ranking, it is also ranked higher in the second ranking. On the other hand, a "discordant pair" involves elements that have a different order between the two rankings. The formula for Kendall's tau involves calculating the difference between the number of concordant pairs and the number of discordant pairs. The resulting value, Kendall's tau coefficient, ranges from -1 to 1. A Kendall's tau coefficient close to 1 suggests a strong agreement between the two rankings, indicating that the order of elements is highly similar. Conversely, a coefficient close to -1 implies significant disagreement, meaning that the two rankings are markedly different. A coefficient of 0 signifies no correlation between the rankings.

\begin{figure}[ht!]
    \centering
    \includegraphics[width=1.0\linewidth]{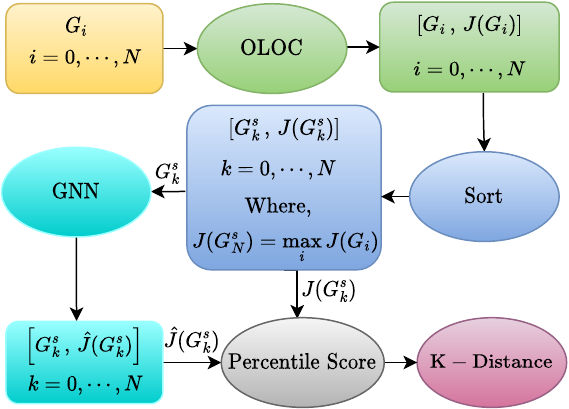}
    \caption{Procedure to calculate Kendall's tau (K) metric.}
    \label{fig:K_metric}
\end{figure}

The procedure to compute K ranking is shown in Fig.~\ref{fig:K_metric}. Suppose, for a population of graphs $G_i$, where $i \in [0, N]$, with N as an integer, we calculate the corresponding values $J_i = J(G_i)$ using the dynamics of the systems and OLOC. These value represent true targets, or labels. If these graphs are sorted based on their $k$ values, we have graphs $G^s_k$, with $k \in [0, N]$, where $G^s_k$ represents the graph sorted at index $k$. In this notation, $k=0$ corresponds to the worst graph, and $k=N$ corresponds to the best graph (corresponding to the highest J value). Also, the corresponding $J$ values of these sorted graphs are denoted as $J^s_k = J(G_k)$. Then, using a trained model, we can estimate the values of the $G^s_k$ graphs. These estimated values can be represented as $\hat{J}_k^s = \hat{J}_k(G^s_k)$. Subsequently, the Kendall's tau of $J^s$ and $\hat{J}_k^s$ can be computed using \textit{Scipy} library.

\subsubsection{Enumeration Metric} 
\label{sec: NOL}
\begin{figure}[ht!]
    \centering
    \includegraphics[width=1.0\linewidth]{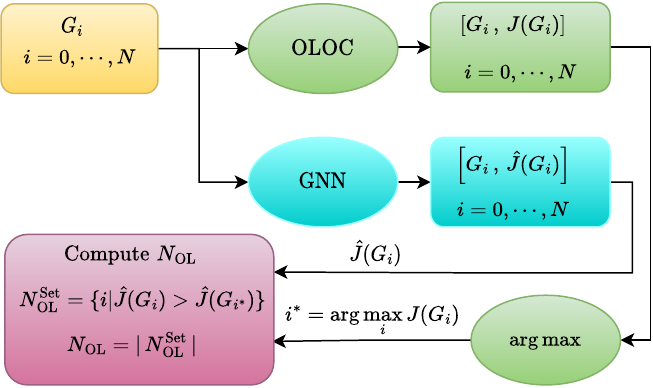}
    \caption{Procedure to calculate $N_{\mathrm{OL}}$ metric.}
    \label{fig:N_OL_metric}
\end{figure}

As previously mentioned, this study focuses on investigating Type 1 problems, described in Section \ref{subSec: Graph_opt}, where we assume that all graphs can be generated, but only some of them can be evaluated within time T. Without utilizing an estimation model, all graphs must be enumerated and evaluated to identify the optimal graph with the best performance. 
When employing the trained model to estimate these graphs and ranking the objective values, an important metric involves determining the number of obtained top-ranked graphs that should be assessed by the OLOC in the next step to guarantee the selection of the optimal graph. The procedure to compute this metric is shown in Fig.~\ref{fig:N_OL_metric}, which involves the following steps:

\begin{enumerate}
    \item Compute $J_i = J(G_i)$ using system dynamics and OLOC analysis for all generated graphs (this includes training and test sets).
    \item For the same graphs, use the trained GNN to estimate the objective value: $\hat{J}_i = \hat{J}(G_i)$.
    \item Compute the optimal graph using the $J$ values and find the corresponding graph that yields the best performance. Denote the index of this optimal graph as $i^{\ast}$ and its corresponding value as $J^{\ast} = J(G_{i^{\ast}})$.
    \item Compute $\hat{J}(G_{i^{\ast}})$ and determine, based on the values of $\hat{J}(G_i)$, how many graphs have higher performance than $\hat{J}(G_{i^{\ast}})$, denoting this count as $N_{\mathrm{OL}}$.
\end{enumerate}

If the variable $N_{\mathrm{OL}}$ is zero, it implies that the estimated optimal graph is indeed the true optimal graph. If it equals $m$, it suggests that there are $m$ graphs mistakenly estimated to have higher performance than the true optimal graph. As a result, to accurately estimate the correct optimal graph, after using GNN to compute the best graphs and ranking them, $m+1$ graphs should be evaluated using the true model (OLOC) to obtain the true optimal result. A smaller value for $m$ is preferable because it reduces the number of graphs that need to be enumerated to obtain the true optimal one.

\subsubsection{Sub-optimality Metric} 
\label{sec: Nsub}

\begin{figure}[ht!]
    \centering
    \includegraphics[width=1.0\linewidth]{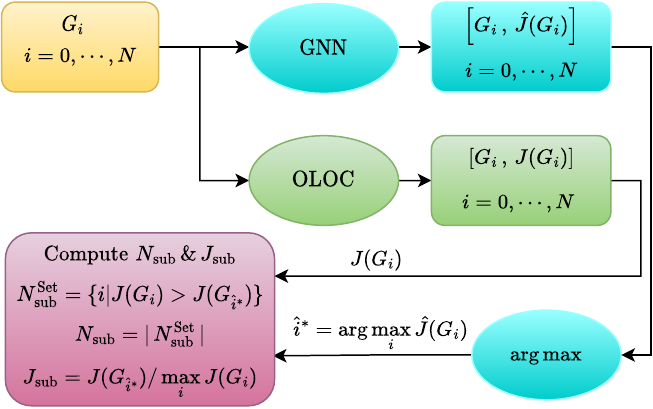}
    \caption{Procedure to calculate $N_{\mathrm{sub}}$ metric.}
    \label{fig:N_sub_metric}
\end{figure}

This metric assesses the performance of the estimated optimal graph. The procedure includes evaluating both the estimated optimal graph, $G_{\hat{i}^{\ast}}$, and its corresponding objective value $\hat{J}^{\ast}=J(G_{\hat{i}^{\ast}})$, in contrast to other objective values ($J(G_i)$) for all graphs in the population, as depicted in Figure ~\ref{fig:N_sub_metric}. As shown in this figure, for each graph $G_i$, both of the GNN model, and system dynamics and OLOC analysis, are employed to compute their respective objective values: $\hat{J}(G_i)$ and $J(G_i)$. Among these N population graphs, the index ($\hat{i}^{\ast}$) that maximizes the $\hat{J}$ values is determined. The J value of this graph index, $J(G_{\hat{i}^{\ast}})$, is then compared to the objective values of all other graphs $J(G_i)$. The resulting metric $N_{\mathrm{sub}}$ indicates the number of graphs that outperform the estimated optimal graph. A value of 0 signifies that the estimated optimal graph is the true optimal graph, and as this value increases, it indicates the sub-optimality of the estimated optimal graph.

Additionally, another variable, denoted as $J_{\mathrm{sub}}=J(G_{\hat{i}^{\ast}})/\max_iJ(G_i)=J(G_{\hat{i}^{\ast}})/J(G_{i^{\ast}})$, is presented. This value signifies the relative importance of the estimated optimal graph's objective value in comparison to that of all other graphs. The rationale behind including this metric is that there may be certain scenarios (heat loads) where the objective values of graphs are nearly identical. In these instances, $N_{\mathrm{sub}}$ may be big because the GNN struggles to precisely estimate the objective value. Consequently, one might perceive this as an unfavorable solution. However, given that many graphs exhibit nearly identical objective values, the specific estimation of the optimal graph becomes less critical. To address this concern, $J_{\mathrm{sub}}$ is also reported. If $J_{\mathrm{sub}}$ approaches 1.0, it indicates that the objective value of the estimated optimal graph closely aligns with the objective value of the true optimal graph.

\subsection{Datasets}
\label{Sec: data}
As previously mentioned, our study specifically addresses Type 1 problems discussed in Section \ref{subSec: Graph_opt}. Consider a case with $n_g$ graphs, each executed under $n_s$ different scenarios (heat loads). The scenarios lead to different node features and objective values. Therefore, the total number of instances, denoted as $N$, equals $n_g \times n_s$. Given the potential for numerous scenarios in engineering systems, $n_s$ tends to be large, resulting in a correspondingly large value for $N$. As an example for the case of a fluid-based thermal management system, each scenario could represent different heat loads for all nodes. As these heat loads are continuous variables, any value is possible, depending on user preferences and the application's heat load requirements. Consequently, it proves valuable to train the GNN on a subset of these scenarios, denoted as $n_s^{\mathrm{train}}$, where $n_s^{\mathrm{train}} << n_s$. The remaining scenarios are labeled as $n_s^{\mathrm{test}} = n - n_s^{\mathrm{train}}$.

As a consequence, within a population of $N$ cases, only $N_{\mathrm{train}} = n_g \times n_s^{\mathrm{train}}$ cases undergo evaluation using the exact model to compute their corresponding objective values. The remaining cases, i.e., $N_{\mathrm{test}} = n_g \times n_s^{\mathrm{test}} = N - N_{\mathrm{train}}$, are not utilized in training but are estimated based on the GNN that has been trained. In this paper, $N_{\mathrm{train}}$ comprises $30\%$ of the data, while $N_{\mathrm{test}}$ constitutes $70\%$ of the data.

\section{Case Study: Fluid-Based Thermal Management System Optimization}
 \label{Sec: HVAC}

This article focuses on a class of fluid base thermal management systems that yield multi-split configurations. We explored these systems in Ref.~\cite{bayat2023multi,bayat2023extracting}, where graph-based models are developed to represent the system configurations and dynamics. We also delved into open-loop optimal control for these systems. In the control problem, we aim to determine the optimal flow rate trajectory for each pipe, maximizing system performance while adhering to component temperature constraints, refer to Figure \ref{fig:whole_system}. The flow rates are regulated by the valves. Here, the system performance is quantified by thermal endurance, meaning our goal is to maximize the duration the device remains operational while ensuring all temperature limits are met.

\begin{figure}[ht!]
    \centering
    \includegraphics[width=1.0\linewidth]{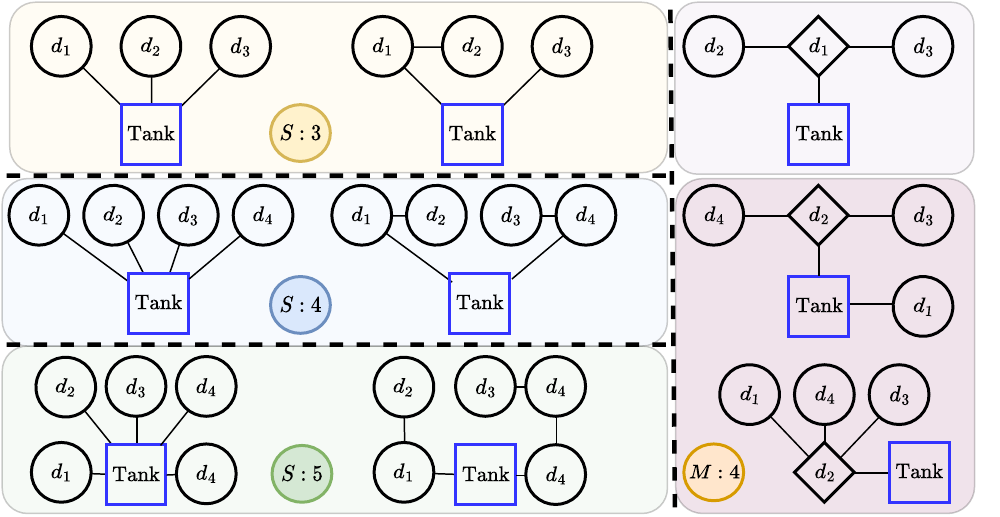}
    \caption{
Illustration presents some possible graphs considered in the training and test data. 'S' denotes a single split, and 'M' refers to multi-split fluid-based thermal management systems.}
    \label{fig:Diff_cases}
\end{figure}

Figure ~\ref{fig:Diff_cases} presents a graphical representation illustrating the modeling of some possible graphs from two distinct system architectures:  multi-split (denoted as $M$) and single-split (denoted as $S$) cases. These scenarios involve the utilization of 3 to 5 CPHXs, each of which contributes a heat load represented by $d_i$ at node $i$. It should be noted that the Liquid-to-Liquid Heat Exchanger (LLHX) nodes are not depicted in this plot, as they carry no heat load and remain consistent across all configurations.

When the number of nodes in the systems increases, we encounter a significant combinatorial explosion in the number of graphs. For instance, transitioning from 3 to 6 nodes results in an increase in the number of possible graphs from 13 to 4051. Consequently, when considering a scenario with 6 nodes, each with a heat load denoted as $d_i$ (for $i$ ranging from 1 to 6), in order to determine the optimal design, one would need to enumerate all 4051 cases and subsequently select the best one. This approach was thoroughly investigated in Ref.~\cite{bayat2023multi}.

Additionally, various machine learning techniques, including deep neural networks, were employed to extract interpretable design knowledge from the dataset derived in  Ref.~\cite{bayat2023multi}. This method proved effective in providing a reliable estimate of good designs, thus eliminating the need to sift through a vast number of cases to find the optimal design. However, the method was limited as when the number of nodes increases, a new model should be trained. 

The studies conducted in Ref.~\cite{bayat2023extracting} and Ref.~\cite{bayat2023multi} revealed insights regarding the optimizations of the active systems under the consideration. It was found that, when provided with the number of CPHXs and their respective heat loads (for example, 4051 for 6 nodes), it is possible to make accurate estimates of configurations with better performances. Additionally, it was noted that the topology of the graph plays a crucial role in making these estimations. For example, when considering three nodes with heat loads denoted as $d_1$, $d_2$, and $d_3$, it was observed that if the heat load values are close to each other, configurations where all three nodes are in parallel tend to perform better \cite{bayat2023extracting}. Conversely, if the heat loads are significantly different, as an example $d_1 >> d_2 >> d3$, a graph with a single branch, where the node with the highest heat load is closest to the tank and the node with the least heat load is farthest from the tank, performs better.

The findings from the aforementioned researches highlight the significance of the graph's topology and loading scenario in estimating the performance of different configurations given specific heat loads \cite{bayat2023multi, bayat2023extracting}. Here we employ GNNs as they take these non-Euclidean relationships into account, making them a suitable choice for estimating the optimal graph design.

\begin{figure}[ht!]
    \centering
    \includegraphics[width=1.0\linewidth]{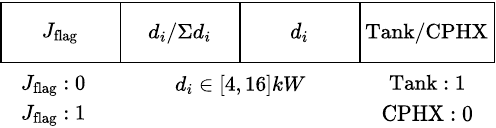}
    \caption{Illustration of node features. The initial feature determines whether the node has a junction, the second feature indicates the relative value of the heat load of that node compared to other nodes, the third feature displays the absolute heat load of the node, and the final feature determines whether the node is a Tank or a CPHX.}
    \label{fig:Node_F}
\end{figure}

To train the GNN, training data was created using the graphs depicted in Fig.~\ref{fig:Diff_cases} while varying the heat loads, where each node's heat load $d_i$ ranged from 4 kW to 16 kW. In order to apply the GNN effectively, node features needed to be defined. As illustrated in Fig~\ref{fig:Node_F}, four features were considered for each node:

\begin{enumerate}
    \item The first feature indicates whether the node has a junction or not.
    \item The second feature represents the relative value of heat load at that node compared to all other nodes.
    \item The third feature reflects the absolute value of heat load at that node.
    \item The fourth feature identifies whether the node is a tank node or a CPHX node.
\end{enumerate}

\begin{figure}[ht!]
    \centering
    \includegraphics[width=1.0\linewidth]{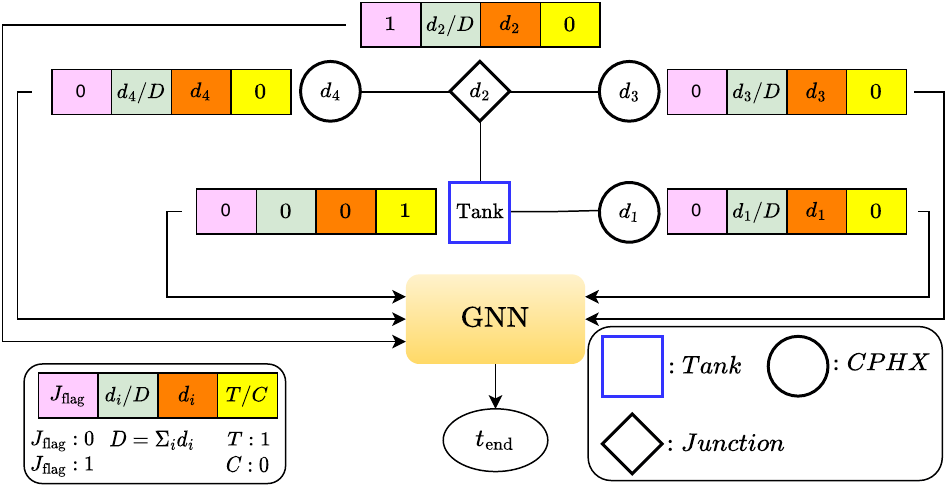}
    \caption{An illustration of the input and output of the Graph Neural Network (GNN) for a graph with a multi-split case featuring a tank and four nodes, one of which is a junction.}
    \label{fig:Graph_Attention}
\end{figure}

In Fig.~\ref{fig:Graph_Attention}, you can observe how the GNN operates. Here, a graph is presented with 4 nodes, where one of them has a junction. The features of all nodes are defined and passed to the GNN model, which estimates the thermal endurance $t_{\mathrm{end}}$. Consequently, in the context of GNN, graph regression is employed to achieve this goal.

\begin{figure}[ht!]
    \centering
    \includegraphics[width=1.0\linewidth]{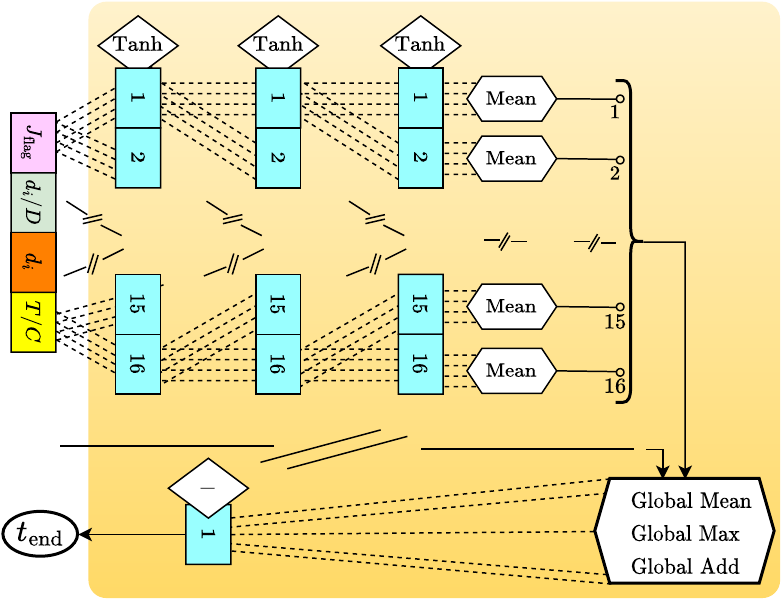}
    \caption{Illustration of the internal layers of the GNN.}
    \label{fig:Graph_All_cropped}
\end{figure}

The GNN utilized for this task is depicted in Fig.~\ref{fig:Graph_All_cropped}. This GNN comprises 3 GAT layers, each with 16 features and a hyperbolic tangent (tanh) activation function. Additionally, each layer includes 4 attention heads. The average over all 4 heads is computed to obtain the node-level features. Subsequently, 3 pooling layers are employed, including Global Mean, Global Max, and Global Add, to transform the node-level features into graph-level features. Finally, a linear layer is used to map these graph embedding features to estimate the thermal endurance.

As outlined in Ref.~\cite{zhou2020graph}, the design of a Graph Neural Network (GNN) involves four essential steps: (1) identifying the graph structure, (2) specifying the graph type and scale, (3) formulating the loss function, and (4) constructing the model using computational modules. In the context of step 1, this study aligns with structural scenarios as the graph structure is explicitly defined. The interconnections among components in the thermal base fluid system studied here illustrate how nodes in the graph are linked. In contrast, non-structural scenarios involve implicit graphs, requiring the initial construction of the graph from the task, such as establishing a fully-connected "word" graph for text \cite{zhou2020graph}. Moving to step 2, the GNN employed in this study is homogeneous, and since input features are not time-dependent, the graph remains static. Furthermore, due to the relatively low number of nodes in the studied graph, its scale is considered small. Concerning step 3, our task operates at the graph-level rather than the node or edge level, as we aim to perform regression based on the entire graph. The setting is supervised, with all labels provided for the training data. In step 4, various modules, including propagation and pooling layers, have been utilized to construct the model.

\section{Results}
\label{Sec: Results}
This section presents the results and assesses the method's performance using various metrics on the training data, test data, and an additional test dataset featuring a higher node count than what was originally not included in the training data. The chosen loss function for this study is MSE loss. In the context of this graph regression problem, the GNN provides mapping between the graph and its corresponding objective function value (thermal endurance). The goal is to minimize the MSE between the estimated objective value and the actual objective value provided by the training data.

\begin{figure}[ht!]
    \centering
    \includegraphics[width=0.8\linewidth]{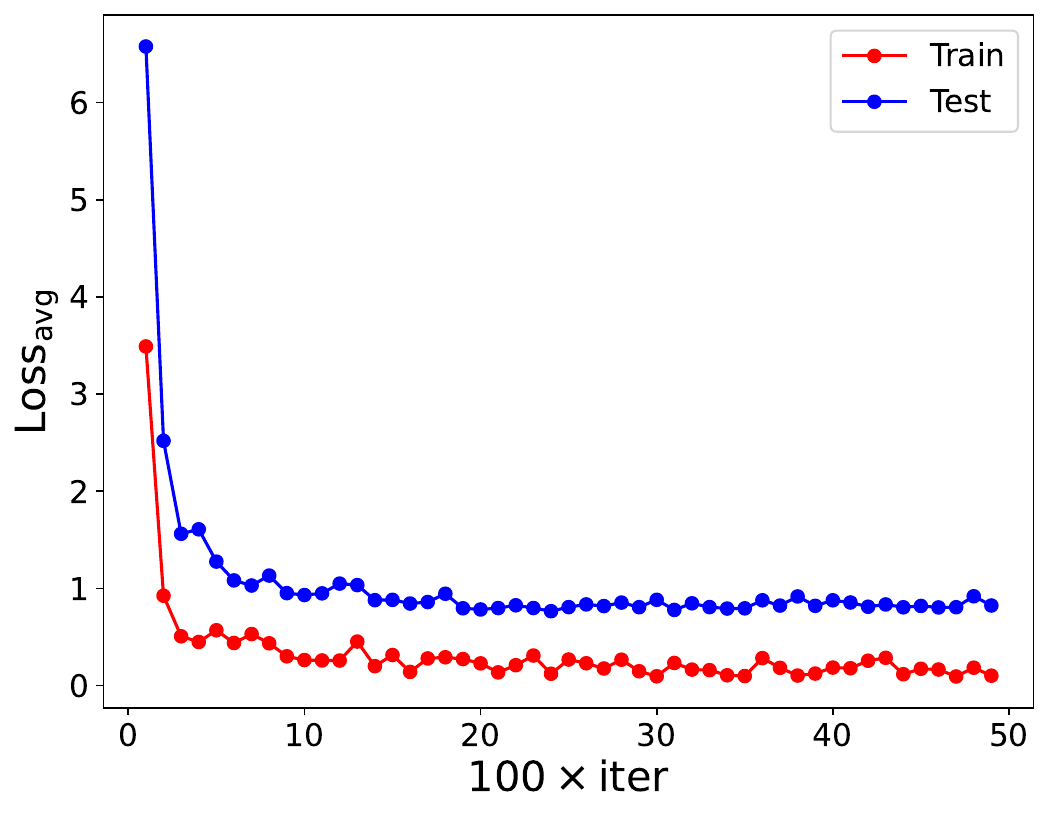}
    \caption{ MSE loss for both training and test data. The x-axis represents every 100 epochs, and the y-axis indicates the average MSE over the corresponding 100 epochs. In total, 50,000 epochs are considered.}
    \label{fig:Loss}
\end{figure}

As mentioned earlier, the objective of this study is to minimize the number of enumerations required to obtain the optimal graph. Consequently, we assumed that the size of the training data should be smaller than that of the test data. The graphs categories depicted in Fig.~\ref{fig:Diff_cases} are examined across different heat loads, resulting in a total population size of 37,329. Thirty percent of the data is allocated for training, while the remaining $70\%$ is reserved for testing. The training process spans 5,000 epochs, utilizing a batch size of 100. Figure~\ref{fig:Loss} illustrates the MSE loss for both the training and test datasets. The x-axis represents every 100 epochs, while the y-axis shows the average value over the corresponding 100 epochs. Observing this figure, it is evident that the model avoids over-fitting. Throughout all epochs, the average loss on both the training and test data exhibits a non-increasing trend. It is worth noting that if the total population size were to increase, the loss would decrease. However, for our specific purpose, the current level of loss is deemed satisfactory as it significantly contributes to reducing the number of enumerations.

\begin{figure}[ht!]
    \centering
    \includegraphics[width=1.0\linewidth]{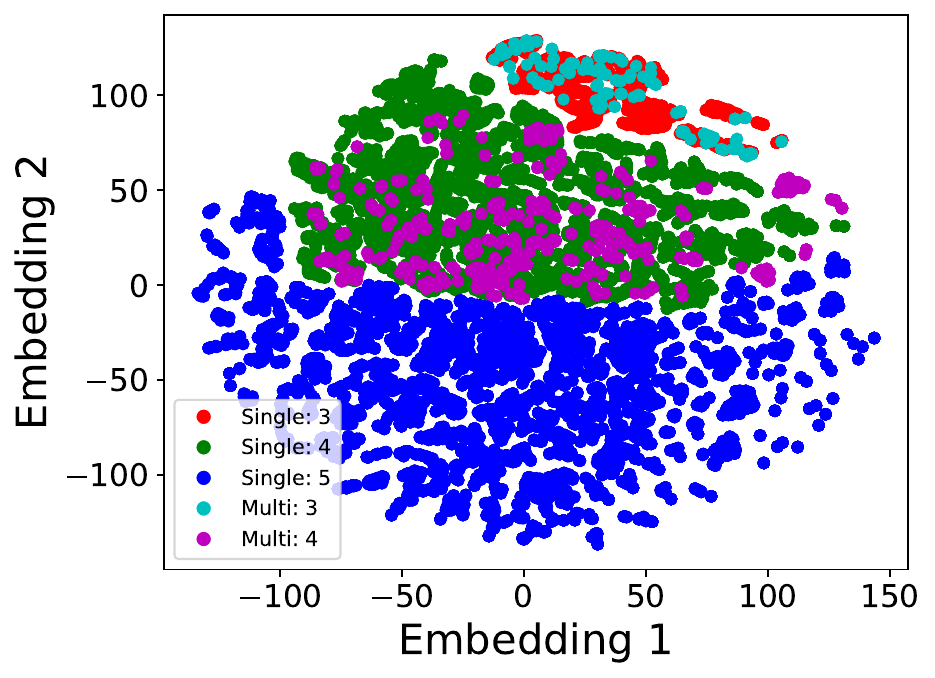}
    \caption{Graph embedding based on T-SNE reduction. Distinct regions in the embedding space correspond to different cases based on the number of nodes. Note that two features are utilized in the image only for visualization purposes. }
    \label{fig:embedding}
\end{figure}

In Fig.~\ref{fig:Graph_All_cropped}, GNN takes the node data of all graph nodes as input and generates an estimate of thermal endurance as output. In the specific architecture illustrated in Fig.~\ref{fig:Graph_All_cropped}, following the global layers (prior to applying the linear layer), the total number of features is 16*3=48. In Fig.~\ref{fig:embedding}, T-SNE \cite{van2008visualizing}, a dimensionality reduction technique, is employed on these 48 features, reducing them to 2 features. This facilitates the visualization of graph embedding in a 2D space. T-SNE, a non-convex tool for visualizing high-dimensional data, achieves this by minimizing Kullback-Leibler divergence.

As depicted in Fig.~\ref{fig:embedding}, distinct regions in the embedding space correspond to different cases based on the number of nodes. It is crucial to emphasize that for visualization purposes, only two features are utilized in the image. Interestingly, using these two features, both single and multiple cases with the same number of nodes are mapped to the same area. However, if more features are normalized, they can be distinguished.

While Graph regression with MSE loss is employed to estimate thermal endurance, the primary objective of this study is to shift from graph enumeration for obtaining the best possible graph with the optimal function value. Here, the aim is to reduce the number of graphs to a subset with a significantly lower number of cases. Subsequently, the graphs within this new set can be enumerated to identify the optimal design. To demonstrate the efficacy of this approach, two metrics are explored in the following subsections.

\subsection{Exploring K Metric}
\label{subSec: Metric1-K}
\begin{figure}[ht!]
    \centering
    \subcaptionbox{}{\includegraphics[width=0.8\linewidth]{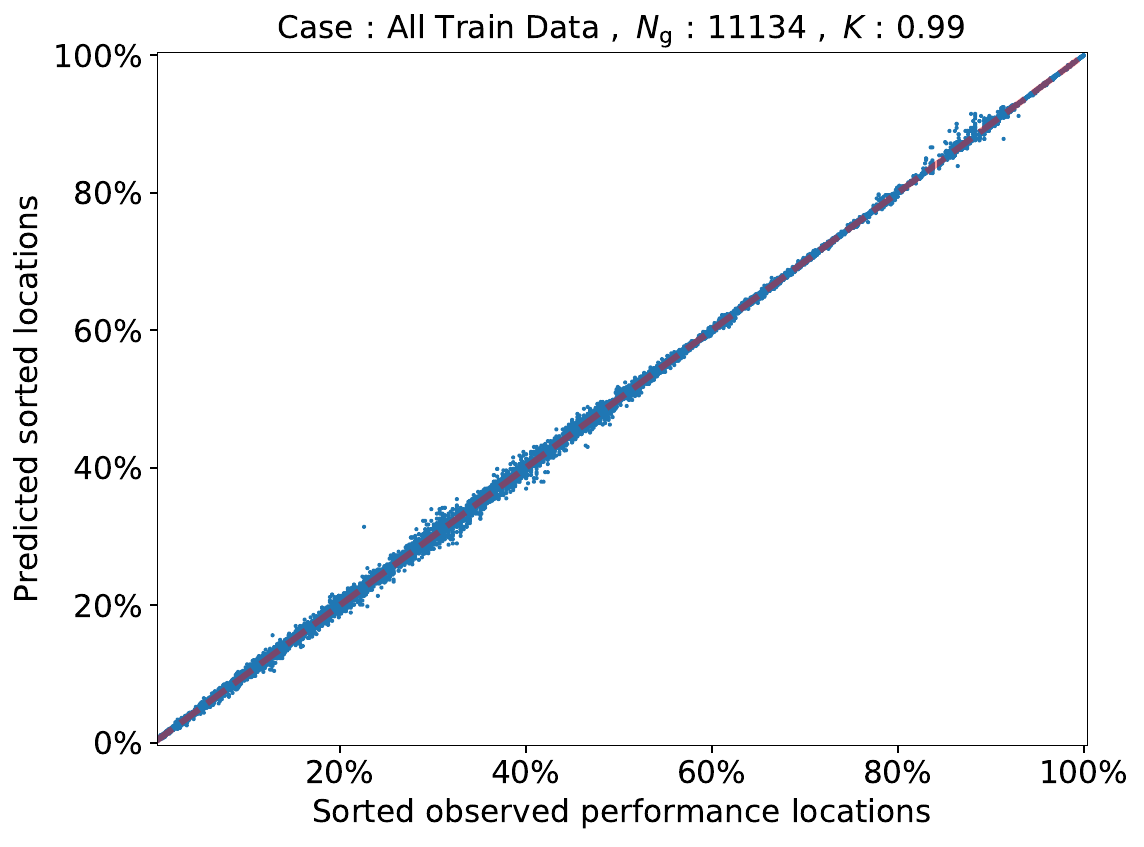}}
    \subcaptionbox{}{\includegraphics[width=0.8\linewidth]{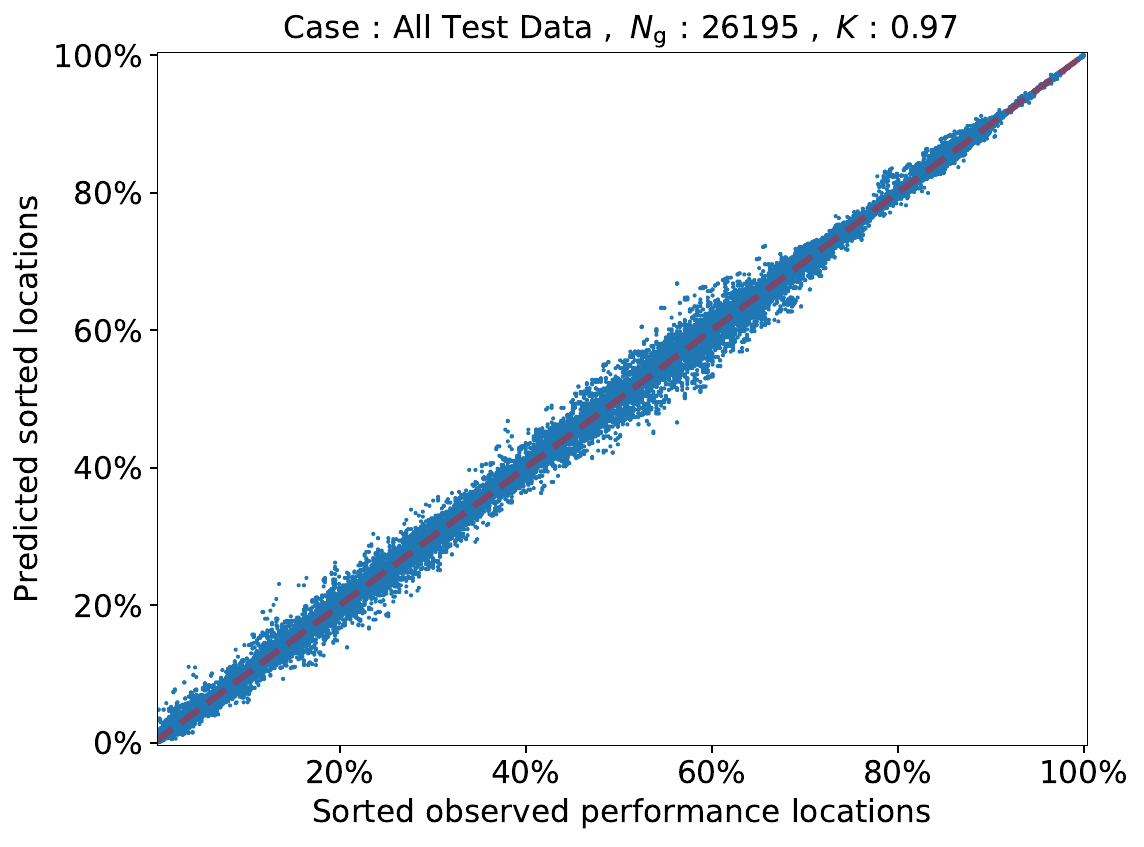}}
    \caption{K metric for both train and test data. $K$ represents the Kendall tau rank distance value, and $N_g$ denotes the total number of graphs. $30\%$ of all data are used for training and $70\%$ for tests.}
    \label{fig:Train_and_Test_K}
\end{figure}

In this section, we present the K metric discussed in Section~\ref{Metric: k} for both training and test data.
In Fig.~\ref{fig:Train_and_Test_K}(a), the x-axis displays the sorted observed performance, while the y-axis indicates the estimated location of those cases. The trained model demonstrates high accuracy on the training data with K-values equal to 0.99. 

Figure~\ref{fig:Train_and_Test_K}(b) depicts a similar plot, this time for the test data. The K factor for the test data is also strong, with a K-value equals to 0.97. This value is obtained for over 26,195 graphs included in the test data. Based on this study, we observe that the model effectively identifies graphs with higher performance, potentially reducing the need for exhaustive enumeration to find the optimal one. This is further demonstrated in the next section using the $N_{\mathrm{OL}}$ and $N_{\mathrm{sub}}$ metrics.

\subsection{Exploring $N_{\mathrm{OL}}$ and $N_{\mathrm{sub}}$ Metrics} 
\label{subSec: Metric1-Ng}

This section delves into the $N_{\mathrm{OL}}$ and $N_{\mathrm{sub}}$ metrics as defined in Sections \ref{sec: NOL} and \ref{sec: Nsub}. The graphs under investigation include single-split cases with 3, 4, and 5 nodes, as well as multi-split cases with 3 and 4 nodes. For a single scenario with prescribed heat loads, the number of different configurations for these graphs is 13, 73, 501, 9, and 88, respectively. For instance, in single-split cases with 5 nodes, given a specific heat load, we anticipate having 501 distinct results.

\begin{figure}[ht!]
    \centering
    {\includegraphics[width=1.0\linewidth]{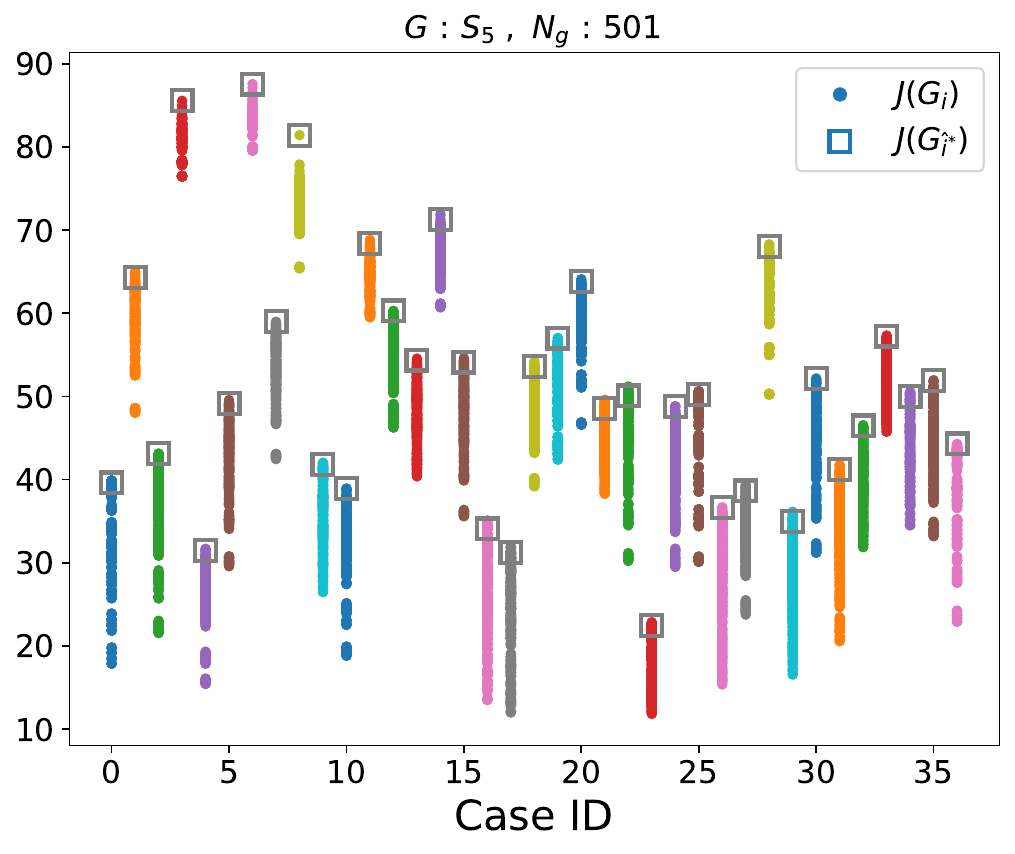}}
    \caption{Comparing the results of OLOC and GNN. A case study involving single-split scenarios with 5 CPHXs under 37 different heat loads. In this context, $J(G_i)$ represents the labels, and $J(G_{\hat{i}^{\ast}})$ indicates the objective value of the estimated optimal graph.}
    \label{fig:N_sub_5}
\end{figure}

To identify the optimal graph in each scenario, for example, in single-split cases with 5 nodes, we would traditionally need to enumerate all 501 cases and select the graph with the highest performance. Alternatively, leveraging the trained GNN allows us to estimate the objective value for each of the mentioned 501 cases. Subsequently, sorting these estimates enables us to identify the index of the graph with the maximum objective function value, denoted as $\hat{i}^{\ast}$. The objective value for this case, $J(G_{\hat{i}^{\ast}})$, can then be compared to the objective values of all the other 501 cases, $J(G_i)$. This comparison is illustrated in Fig.~\ref{fig:N_sub_5} for the single-split case with 5 nodes. For a comprehensive view, please refer to the appendix, Fig.~\ref{fig:Pred_1}, which provides the same plot for all other graph cases with different number of nodes and split types.

The x-axis in Fig.~\ref{fig:N_sub_5} represents 37 different scenarios (different heat loads) for the single-split graph with 5 nodes. For each scenario, there are 501 different graphs, and the y-axis displays the objective values of these graphs. Since the objective function of the system optimizer aimed to increase thermal endurance, a higher value on the y-axis corresponds to better performance, with the topmost point indicating the optimal performance for the given heat load. In this plot, the dot points depict the true objective values for all 501 cases ($J(G_i)$), while the square node represents $J(G_{\hat{i}^{\ast}})$, which shows the objective value of the estimated optimal graph. It is evident that in all 37 cases, the estimated optimal graph is remarkably close to the true optimal one, providing an accurate estimate of the optimal graph.

\begin{figure}[ht!]
    \centering
    {\includegraphics[width=1.0\linewidth]{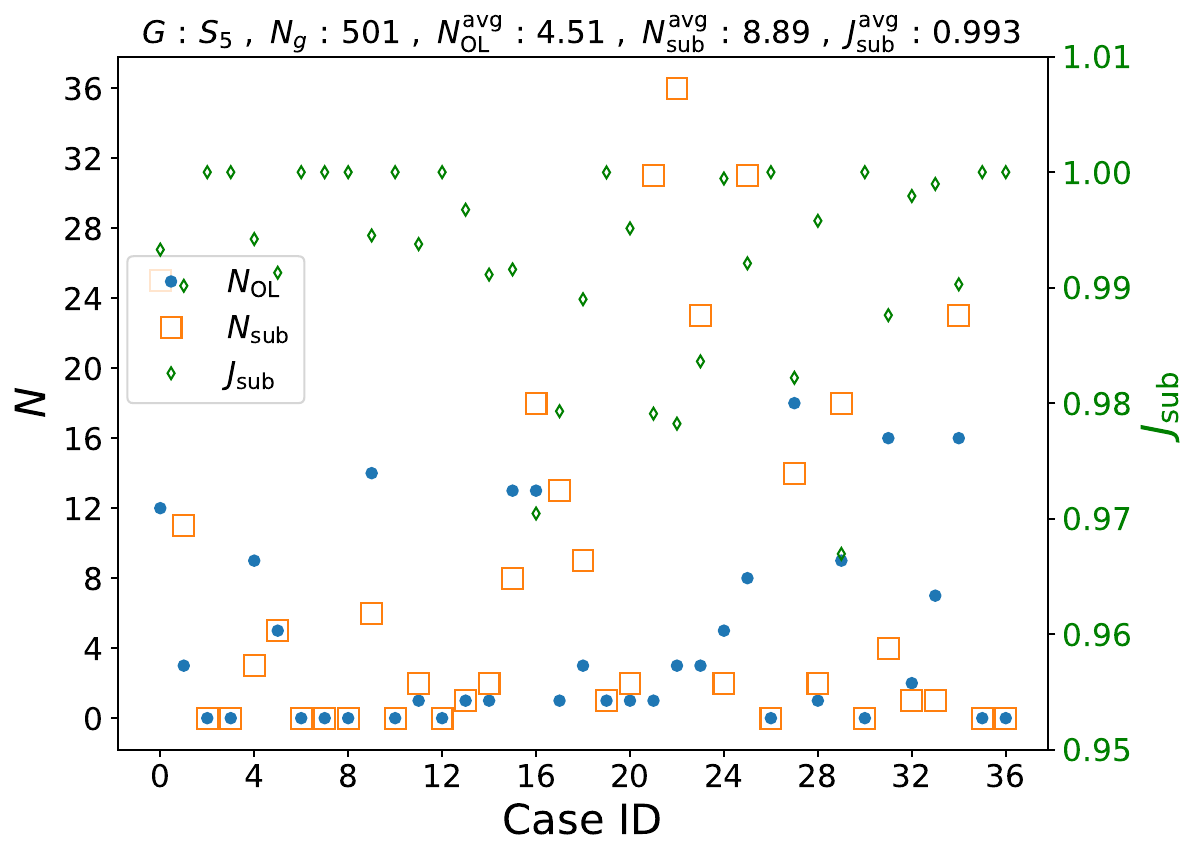}}
    \caption{Comparing the results of OLOC and GNN. A case study on single-split scenarios with 5 CPHXs under 37 different heat loads. In this context, $J(G_i)$ represents the labels, and $J(G_{\hat{i}^{\ast}})$ indicates the objective value of the estimated optimal graph. The figure also shows metrics $N_{\mathrm{OL}}$, $N_{\mathrm{sub}}$, and $J_{\mathrm{sub}}$.}
    \label{fig:Pred_2_5}
\end{figure}

Figure~\ref{fig:Pred_2_5} presents $N_{\mathrm{OL}}$, $N_{\mathrm{sub}}$, and $J_{\mathrm{sub}}$ for the single-split case with 5 nodes. The x-axis aligns with the x-axis in Fig.~\ref{fig:N_sub_5}. The right y-axis displays the relative value of the objective value of the estimated optimal graph, $J(G_{\hat{i}^{\ast}})$, with respect to the correct optimal graph value $J^{\ast}=J(G_{i^{\ast}})$. Here, $J^{\ast}$ is equivalent to $\max_i J(G_i)$ and the $J_{\mathrm{sub}}$ metric was explained in Fig.~\ref{fig:N_sub_metric}. The observed relative value falls within the range of 0.95 to 1, indicating that the objective value of the estimated optimal graph is very close to the objective value of the true optimal graph.

On the left y-axis, the values for $N_{\mathrm{OL}}$ and $N_{\mathrm{sub}}$ are displayed. Notably, the maximum value for $N_{\mathrm{sub}}$ is 36, suggesting that, in the worst-case scenario, only 36 out of 501 cases surpass the estimated optimal graph. Consequently, the estimated optimal graph ranks in the top $8\%$ of the 501 cases, and the relative value of the objective function for this case, with respect to the optimal solution, exceeds 0.95 (depicted on the right y-axis). Therefore, the estimated optimal graph is very close to the optimal solution in the objective function value space.

Furthermore, as shown in Fig.~\ref{fig:Pred_2_5} the maximum value for $N_{\mathrm{OL}}$ is 18, signifying that to identify the optimal graph, only the top 18+1 cases obtained from the GNN need to be enumerated and evaluated by OLOC. Consequently, instead of enumerating all 501 cases, only 19 cases need to be considered to pinpoint the optimal design. The figure's title displays the average values of $N_{\mathrm{OL}}$, $N_{\mathrm{sub}}$, and $J_{\mathrm{sub}}$ over 37 scenarios (different loads). Notably, the average value for $N_{\mathrm{OL}}$ is 4.51, for $N_{\mathrm{sub}}$ is 8.89, and for $J_{\mathrm{sub}}$ is 0.993. This implies that, on average, only 5 enumerations are required to achieve an optimal solution. Furthermore, the estimated optimal solution, on average, holds the 9th rank out of 501 possible ranks and its objective value comparing to the true optimal value is 0.993.

\subsection{Testing the Proposed Workflow on Systems with Increased Complexity}
\label{subSec: result_test}
So far, our population comprises of single-split cases with 3 to 5 nodes and multi-split cases with 3 to 4 nodes. These cases were subjected to various heat loads, and subsequently, we partitioned the total population into training and test sets to analyze the outcomes. An intriguing aspect of our study involves investigating the performance of the trained model on new graphs with higher complexity—graphs not included in the training or test sets. This introduces a novel challenge to the system, prompting us to explore whether the Graph Neural Network (GNN), leveraging its message-passing units, can provide accurate estimations. To explore this, we focused on a single-split case with 6 nodes, corresponding to 4051 distinct graphs. Enumerating all 4051 cases to obtain the optimal solution for a given heat load is computationally demanding. Employing 64 workers in parallel would take approximately 12 hours to evaluate all these cases. However, leveraging the trained GNN allows us to accomplish this task in a matter of seconds. The computation expenses for all presented findings were obtained using a workstation featuring an AMD EPYC 7502 32-Core Processor @ 2.5 GHz, 64 GB DDR4-3200 RAM, LINUX Ubuntu 20.04.1, and Python 3.8.10.

\begin{figure}[ht!]  
    \centering
    \subcaptionbox{}{\includegraphics[width=0.8\linewidth]{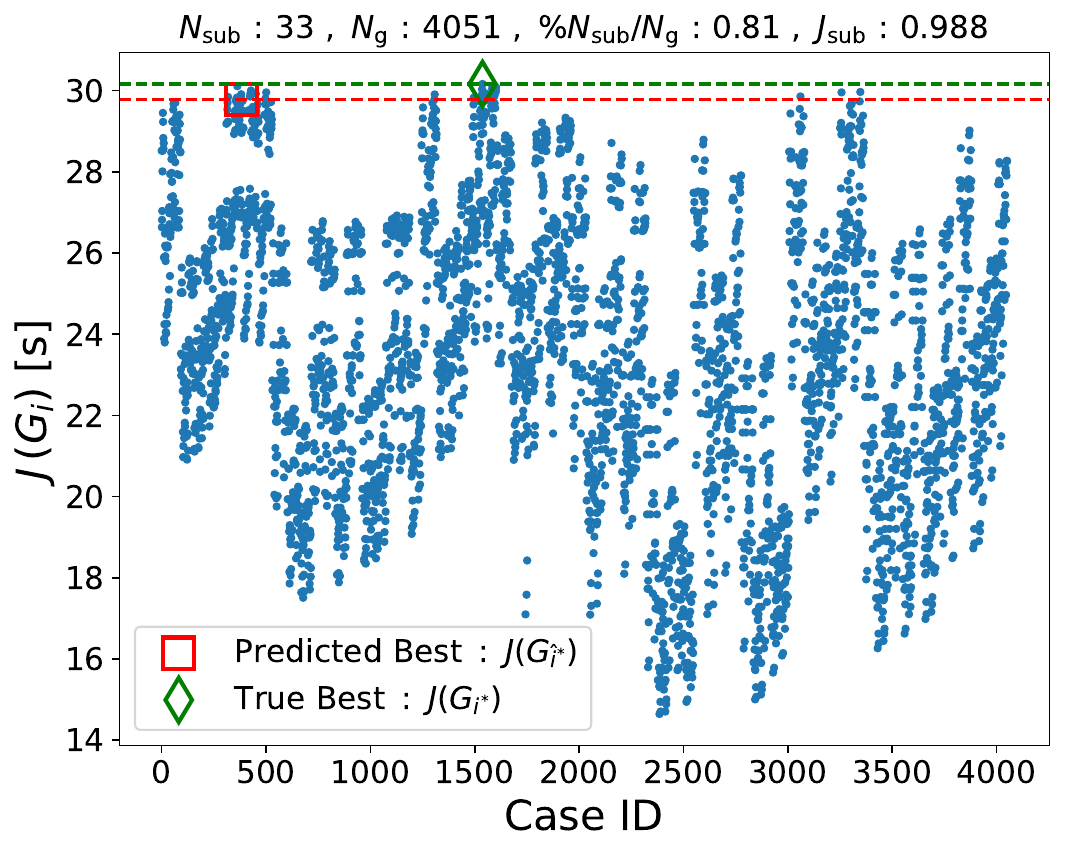}}
    \subcaptionbox{}{\includegraphics[width=0.8\linewidth]{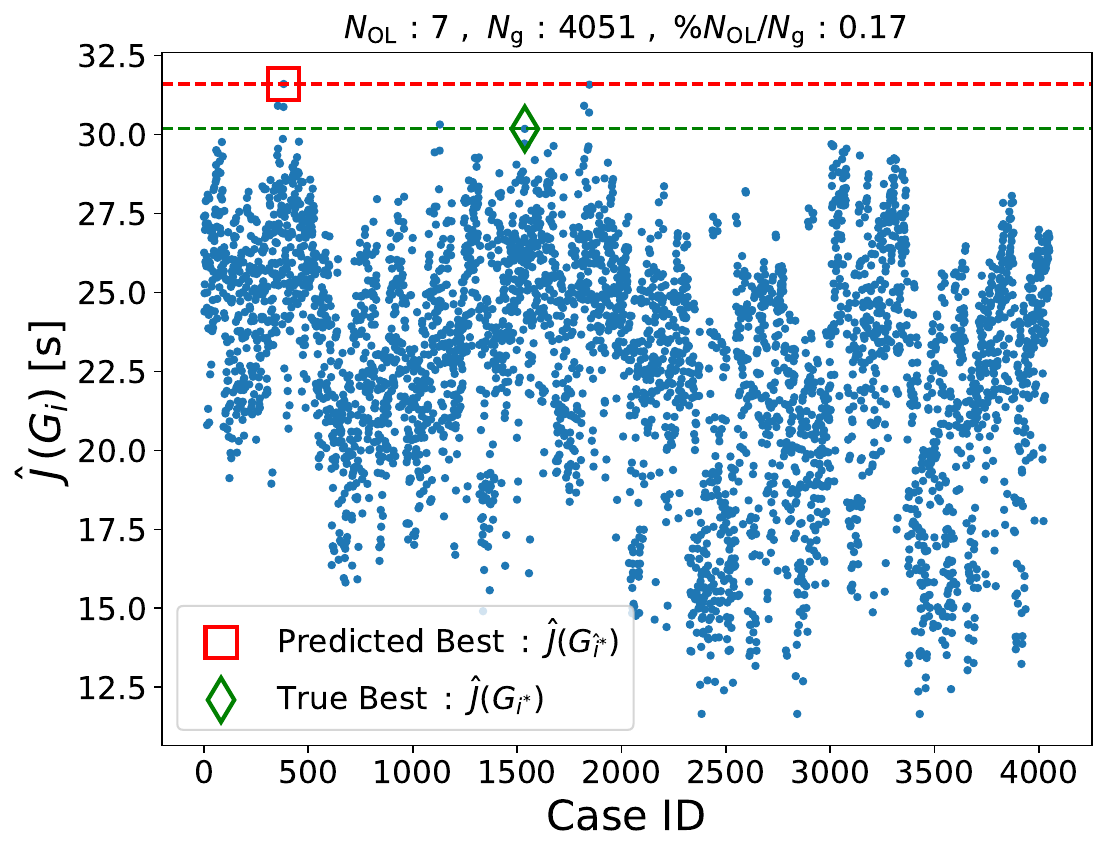}}
    \caption{Examining the results of the trained Graph Neural Network (GNN) on a new, more complex graph that was not included in the training data, and calculate different metrics. This new graph has a single-split structure with 6 CPHXs, encompassing 4051 different graphs given a heat load. Here, $J(G_i)$ represents the labels, and $J(G_{\hat{i}^{\ast}})$ indicates the objective value of the estimated optimal graph.}
    \label{fig:S6}
\end{figure}

The findings of this GNN study are illustrated in Fig~.\ref{fig:S6}. In both subplots, the x-axis represents the indices of the 4051 cases, where the applied heat load is $[12, 10, 8, 6, 5, 4]$ kW. In subplot (a), the y-axis displays the corresponding objective values. The predicted best solution, denoted by a square, represents $J(G_{\hat{i}^{\ast}})$, where $\hat{i}^{\ast}$ is the estimated optimal graph index obtained by GNN. The True Best is indicated by $J(G_{i^{\ast}})$. The title of the image includes additional information on the evaluation metrics. As an example $N_{\text{sub}}$ is reported as 33, indicating that the estimated optimal solution ranks 33rd. The objective value of the predicted solution to the best solution is 0.988, highlighting the proximity of the predicted solution (ranking 33) to the optimal solution.

In Fig~.\ref{fig:S6}(b), the estimated objective values of the same cases are depicted. Here, $N_{\mathrm{OL}}$ is 7, signifying that only 7 enumerations are needed to obtain the true optimal solution. Consequently, the number of populations that need to be enumerated to achieve the optimal solution decreases from 4051 to 7, constituting only $0.17\%$ of the total population. This study demonstrates that the trained model performs well on new, more complex graphs not encountered during training. The success is attributed to the GNN's incorporation of message passing through different nodes and utilization of graph topology to estimate the optimal design.

\begin{figure}[ht!]
    \centering
    {\includegraphics[width=0.8\linewidth]{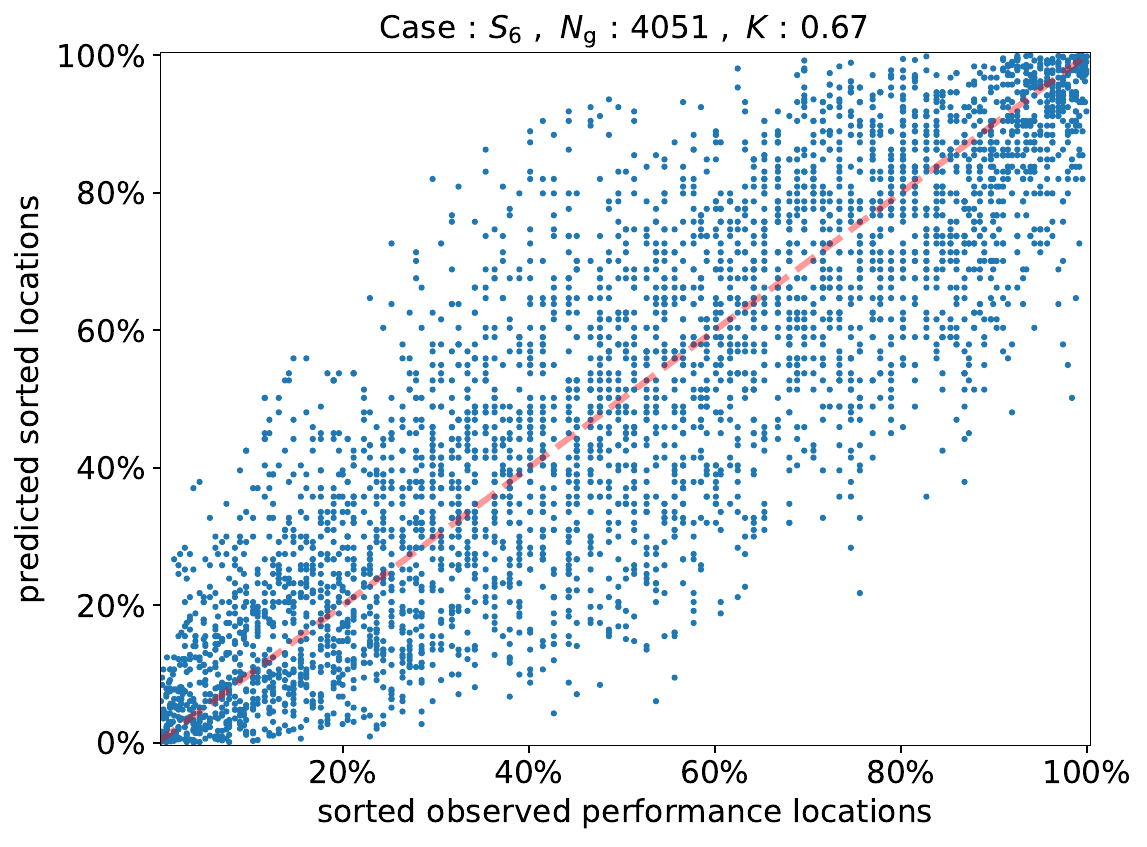}}
    \caption{K-metric result for a new, more complex graph that was not included in the training data. This new graph has a single-split structure with 6 CPHXs, encompassing 4051 different graphs given a heat load.}
    \label{fig:S6_k}
\end{figure}

The K metric for these 4051 cases is depicted in Fig.~\ref{fig:S6_k}. Notably, the metric registers a value of 0.67, which is interesting considering the entirely new graph. Additionally, for both higher and lower objective values, the data exhibits less deviation from the ideal line. This implies that the corresponding K value on graphs with higher objective values (indicating better performance) has more consistency than the cases in the middle, which are distant from the optimal design. Consequently, the model can predict good solutions more accurately than estimating solutions in the middle range. This observation may be attributed to the fact that, in the middle range, there is more data, resulting in greater diversity in the results compared to the upper range.

\section{Limitations and Future Work}
\label{Sec: lim}
While the method discussed in this paper has demonstrated promising results, it is not without limitations. One of the primary constraints lies in the treatment of heat load. In the presented case studies, the heat load of nodes remains constant, indicating no temporal variation. If temporal variations are introduced, employing more intricate GNN models with memory components becomes necessary. Another limitation pertains to the complexity of the optimal control problem. The generation of training data involves solving various OLOC problems, where components such as system dynamics and constraints play pivotal roles. If the dynamics are more intricate or the constraints are more complex, the number of node features could potentially increase. Additionally, node feature selection becomes more critical, necessitating pre-processing to obtain the optimal features for graph regression. These aspects will be addressed in future work. 

Furthermore, as previously stated, this research centers on Type 1 problems, refer to Section \ref{subSec: Graph_opt} where all graphs can be generated within the given time frame T. Nevertheless, there are numerous applications where even generating a graph is computationally expensive. For such problems, classified as Type 2, new methods must be developed. These limitations should be further investigated in future research.

\section{Conclusion}
\label{Sec: conc}

In conclusion, this paper introduces a framework that utilizes Graph Neural Network (GNN) for the graph regression task in the optimization of active engineering systems. By leveraging the GNN-based framework, we address the limitations of traditional optimization methods that require the enumeration of all system configurations and the exhaustive evaluation of each configuration using the computationally expensive OLOC. Our proposed framework eliminates the need for exhaustive evaluation by estimating the objective value of graphs using the trained GNN. The GNN model is trained on a small subset of graphs, and once trained, it can efficiently evaluate the objective values of all enumerated graphs. In our case study, we trained the GNN model using 30\% of the data. 

The results demonstrate the effectiveness of our approach. On a diverse test dataset of 4051 graphs, only the top 7 ranked members identified by the GNN need to be evaluated by OLOC to obtain the true optimal solution. This reduction in the number of enumerations results in a significant decrease of over 99\% in computational effort. On average, the reduction is approximately 92\% for the cases examined in this research, highlighting the efficiency of our approach. Furthermore, the use of OLOC to generate the training data contributes to building GNN knowledge based on optimal solutions. This not only benefits the design of new systems with limited or no design heritage but also provides opportunities for uncovering radically new designs that were not previously considered.

Overall, our GNN-based framework offers a promising approach for optimizing active engineering systems, providing significant computational savings and enabling the exploration of new design possibilities. Future research can focus on expanding the framework to other domains and investigating its scalability and applicability in real-world engineering scenarios.

\appendix   

\section{Performance of the GNN on all graphs}

\begin{figure*}[ht!]
    \centering
    \subcaptionbox{}{\includegraphics[width=0.33\linewidth]{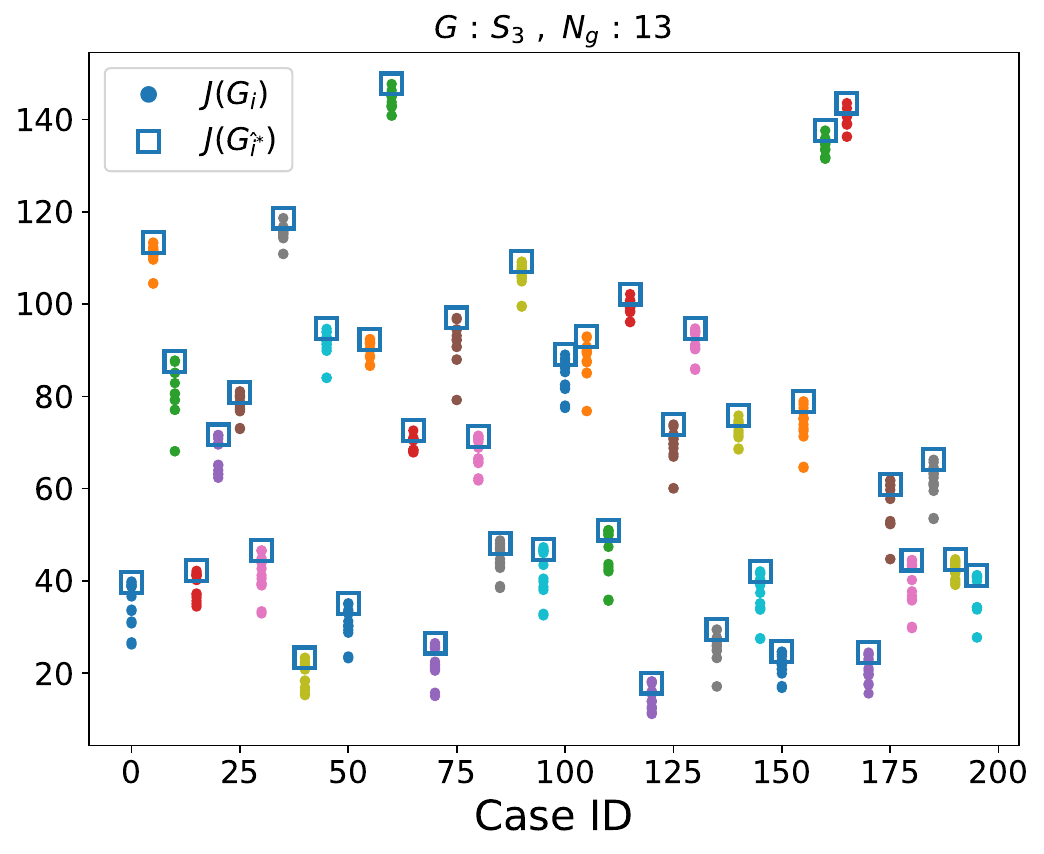}}
    \subcaptionbox{}{\includegraphics[width=0.33\linewidth]{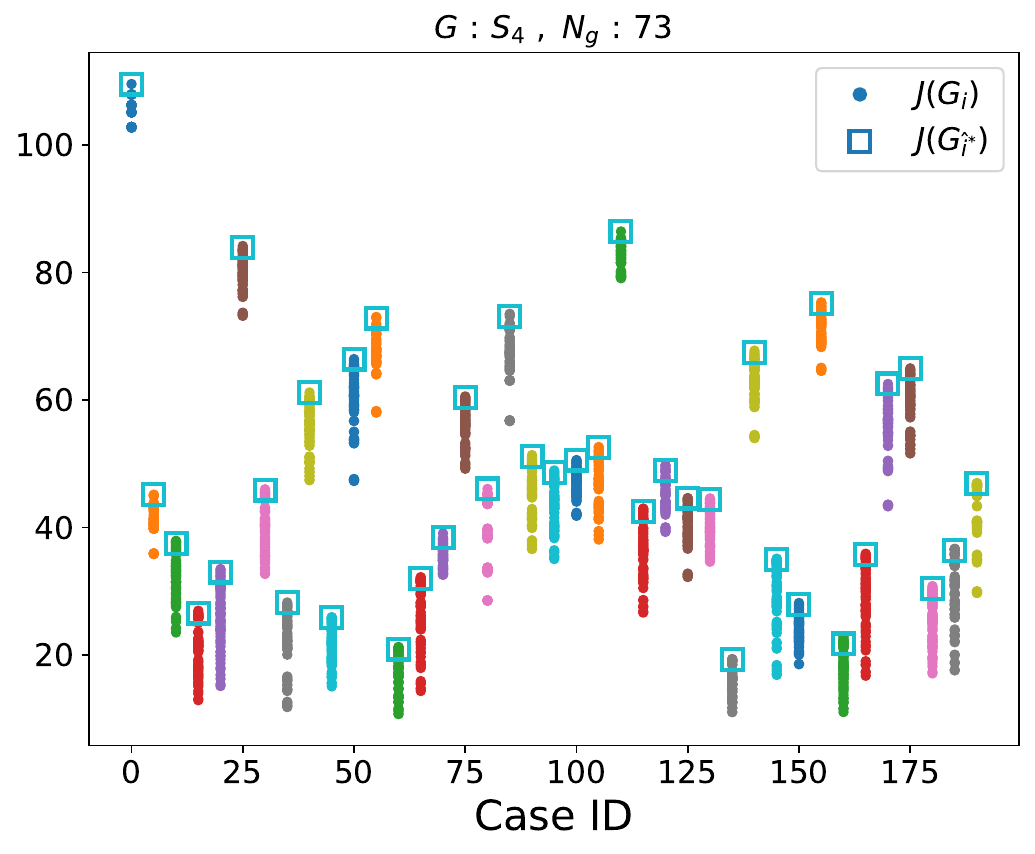}}
    {\includegraphics[width=0.33\linewidth]{Images/all_pop_5_labels_pred.pdf}}
    {\includegraphics[width=0.33\linewidth]{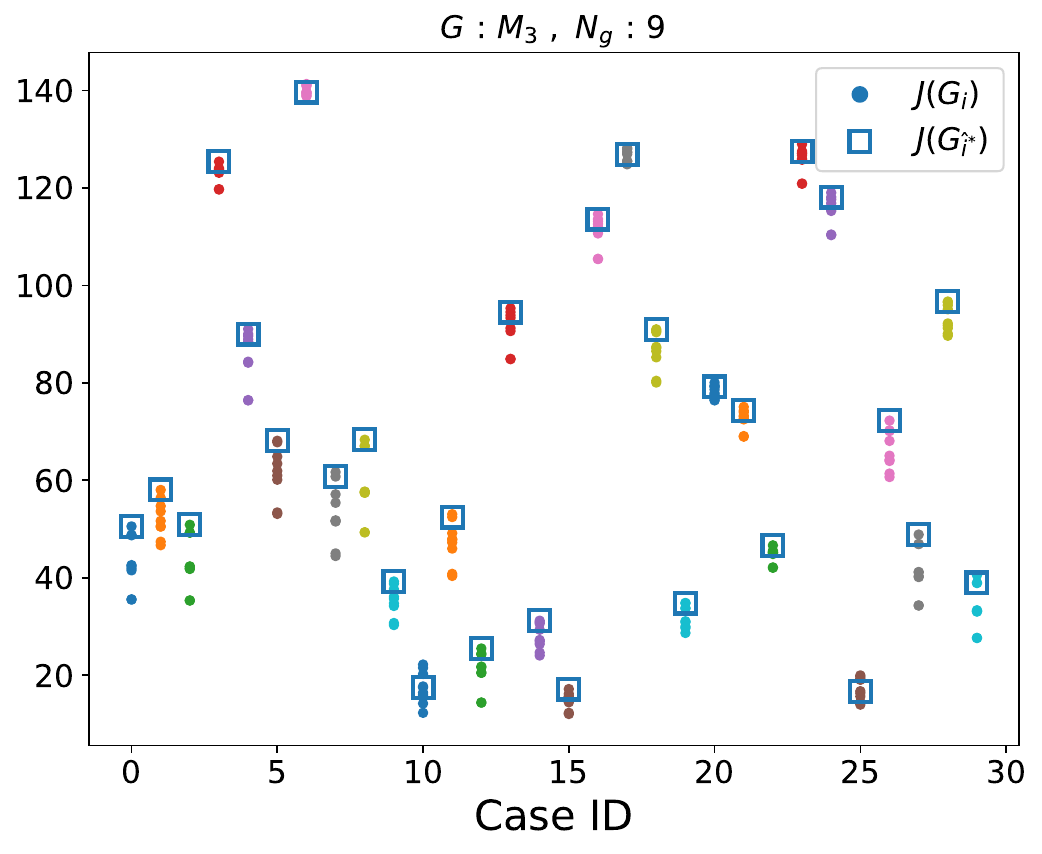}}
    {\includegraphics[width=0.33\linewidth]{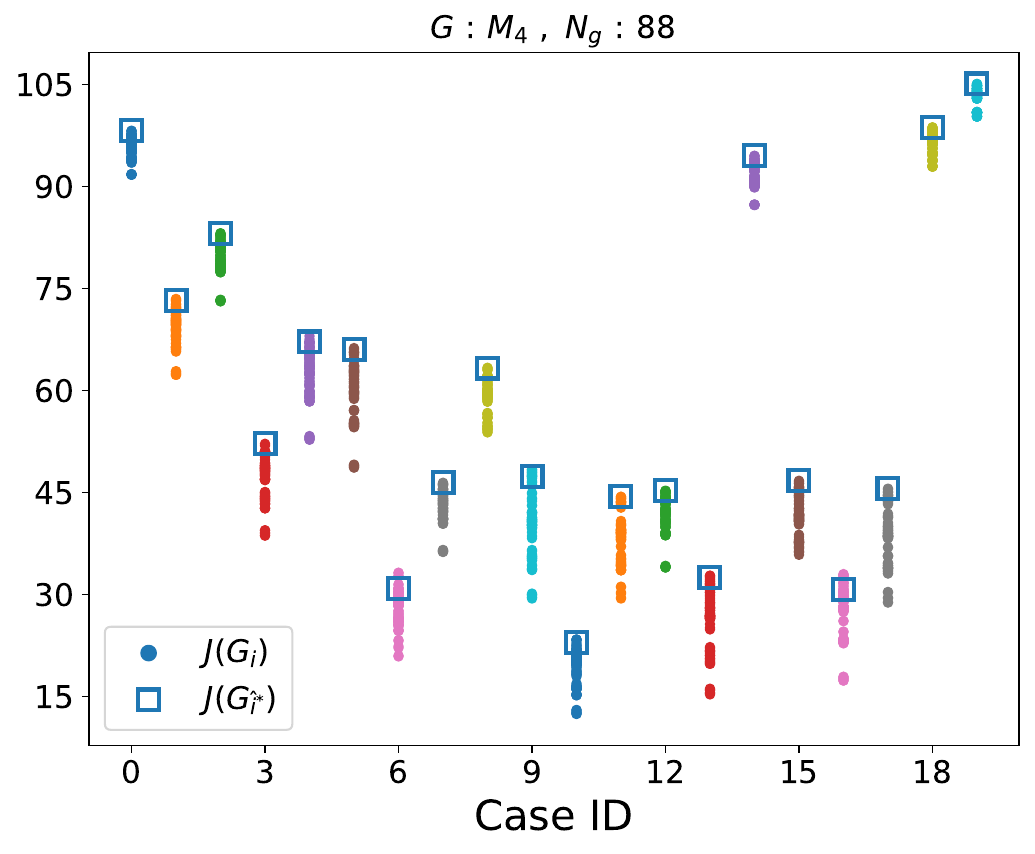}}
    \caption{Comparing the results of OLOC and GNN on both the training and test data. In this context, $J(G_i)$ represents the labels, and $J(G_{\hat{i}^{\ast}})$ indicates the objective value of the estimated optimal graph.}
    \label{fig:Pred_1}
\end{figure*}

\begin{figure*}[ht!]
    \centering
    \subcaptionbox{}{\includegraphics[width=0.33\linewidth]{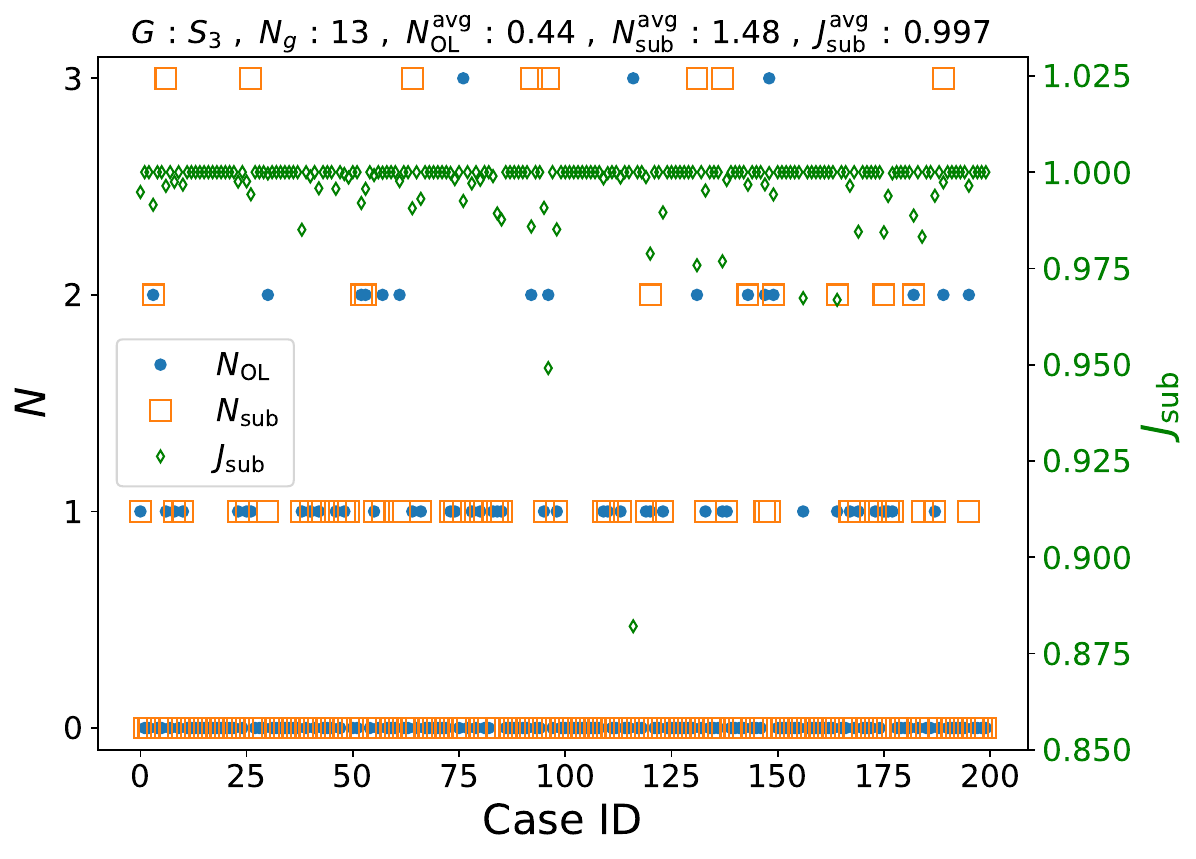}}
    \subcaptionbox{}{\includegraphics[width=0.33\linewidth]{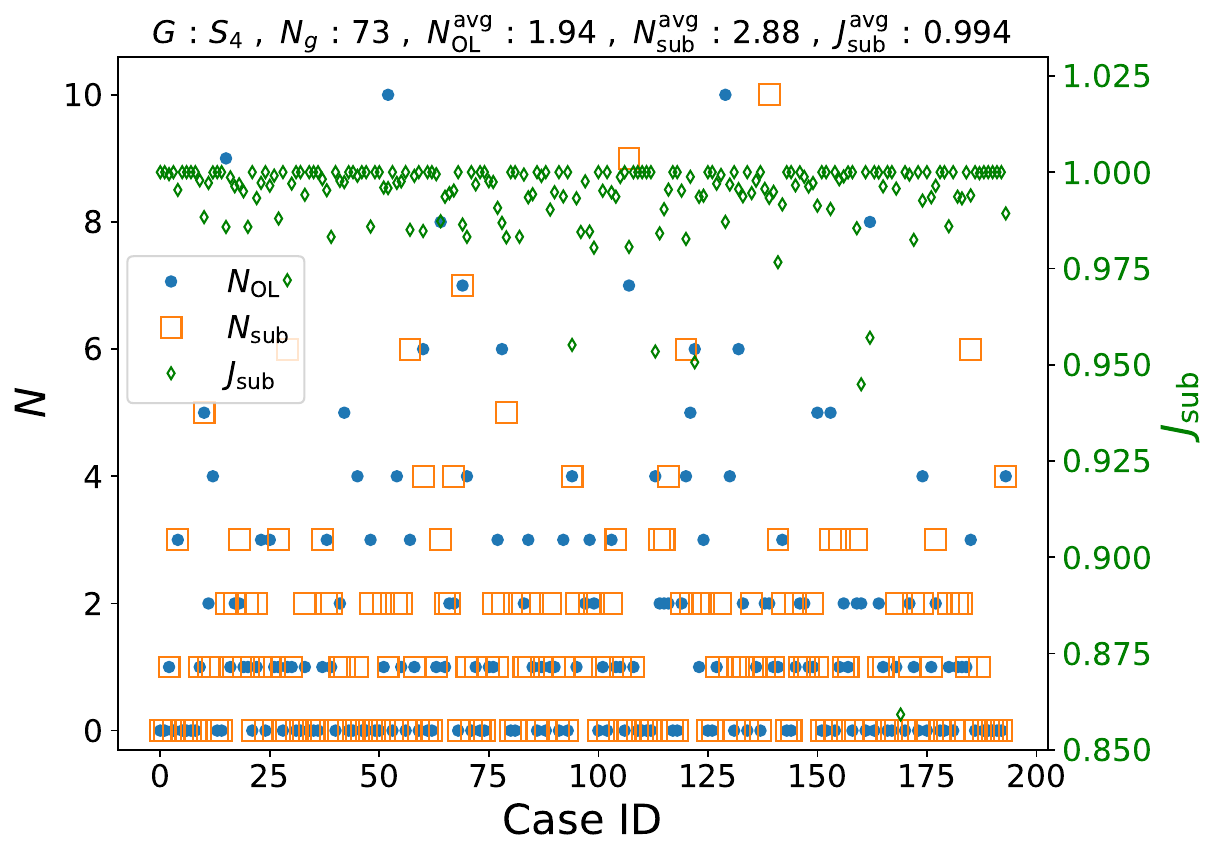}}
    {\includegraphics[width=0.33\linewidth]{Images/N_tg_Combined_S5.pdf}}
    {\includegraphics[width=0.33\linewidth]{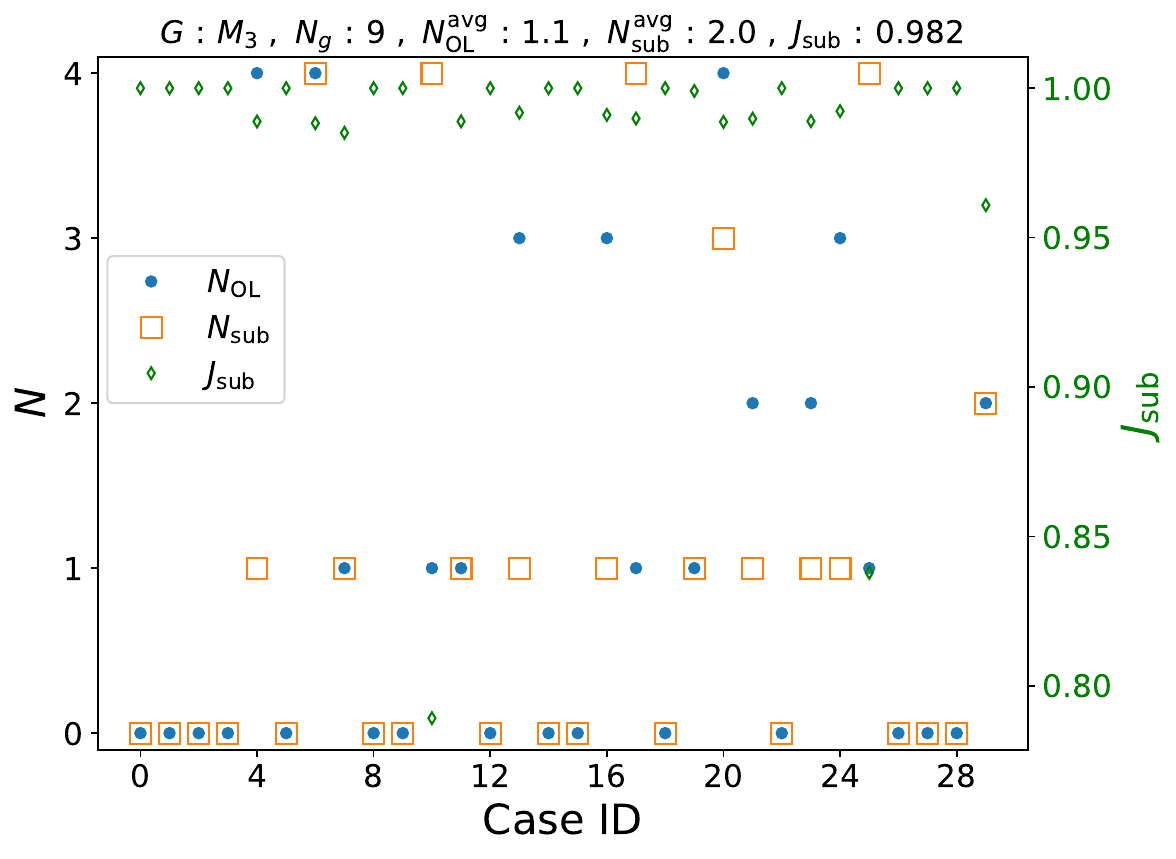}}
    {\includegraphics[width=0.33\linewidth]{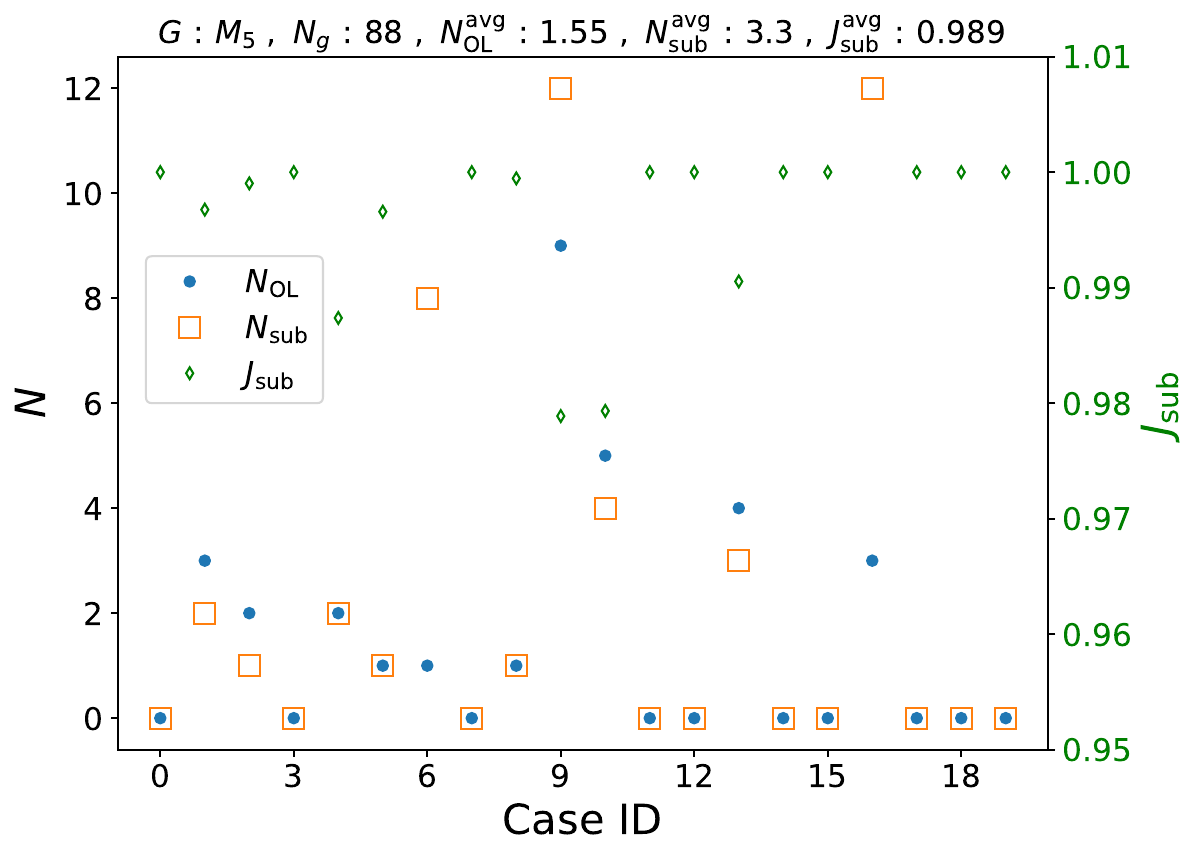}}
    \caption{Comparing the results of OLOC and GNN on both the training and test data. In this context, $J(G_i)$ represents the labels, and $J(G_{\hat{i}^{\ast}})$ indicates the objective value of the estimated optimal graph. Here some metrics such as $N_{\mathrm{OL}}$, $N_{\mathrm{sub}}$, and $J_{\mathrm{sub}}$ are displayed.}
    \label{fig:Pred_2}
\end{figure*}

Figures~\ref{fig:Pred_1} and \ref{fig:Pred_2} present the outcomes for all single-split cases with 3, 4, and 5 nodes, as well as multi-split cases with 3 and 4 nodes.

\bibliographystyle{asmejour}   

\bibliography{Main} 



\end{document}